\documentclass{aa}
\usepackage{graphicx}
\usepackage{natbib}
\usepackage{aalongtable}
\newcommand{\td}{{$\bigtriangleup$}}
\newcommand{\tu}{{$\bigtriangledown$}}
\begin{document}
   \titlerunning{Periodic clustering of exposure ages of iron
meteorites}
   \title{On the periodic clustering of cosmic ray \\exposure ages of iron
meteorites}

   \author{Knud Jahnke}

   \institute{Astrophysikalisches Institut Potsdam, An der Sternwarte
16, D-14482 Potsdam, \email{kjahnke@aip.de}
             }

   \date{Received someday in 2005; accepted later in 2005}

   \abstract{
Two recent papers claimed to have found a periodic variation of the
galactic cosmic ray (CR) flux over the last 1--2~Gyr, using the CR
exposure ages of iron meteorites. This was attributed to higher CR
flux during the passage of the Earth through the spiral arms of the
Milky Way, as suggested by models. The derived period was
143$\pm$10~Myrs. We perform a more detailed analysis of the CR
exposure ages on the same data set, using extensive simulation to
estimate the influence of different error sources on the significance
of the periodicity signal. We find no evidence for significant
clustering of the CR exposure ages at a 143~Myr period nor
for any other period between 100 and 250~Myrs. Rather, we find the
data to be consistent with being drawn from a uniform distribution of
CR exposure ages. The different conclusion of the original studies is
due to their neglecting the influence of (i) data treatment on the
statistics, (ii) several error sources, and (iii) number statistics.
   }

   \maketitle
%

\section{Introduction}\label{sec:intro}
In recent years several authors have looked at a possible influence of
cosmic rays (CRs) on climate, in particular their possible correlation
with cloud cover \citep[e.g.,][]{sven98}. The evidence for such a
correlation, and the question whether CRs influence climate, remains
controversial \citep{kris02,laut03}.

A recent series of publications \citep[henceforth `S02', `S03', and
`SV03']{shav02,shav03,shav03b} presented a model for CR production in
the spiral arms of the Milky Way. There the CR diffusion to earth
during the passage of the solar system through the spiral arms in the
past $\sim$1--2~Gyr was modelled. One of their basic claims was the
existence of a periodic modulation of the CR flux, and a temporal
correlation between the galactic CR influx to earth from their model
predictions and the times of glacial periods on Earth. Now some
authors have begun using these results in further research
\citep{fuen04,wall04,gies05}, while others have challenged the data
handling and significance of the results \citep{rahm04}.

S02/S03/SV03 based the timing of galactic CR peaks in their model on
apparent age clusters of meteorites, found by exposure age dating of
iron meteorites in S02 and S03. However, both papers lack a discussion
of possible sources of error for the statistical significance of their
results. Further, details of the meteorite data treatment are missing
at several points. This prompted us to examine in more detail the
statistical basis of the claimed periodic clustering of CR exposure
ages.
 
In this article we identify several sources of influence on a signal
for a non-uniform distribution of CR exposure ages, and assess their
quantitative strenths. We use the original data used by S03, and
critically follow the analysis methods described in S02/S03. Our aim
is to reevaluate the statistical significance of the S02/S03 results
with respect to CR exposure ages {\em without a priori assumptions}.

\subsection{Cosmic ray exposure ages}

In S02/S03 a connection was drawn between a model of galactic cosmic
ray diffusion put forward by the authors and observational constraints
on absolute timing by CR exposure ages of iron meteorites. The authors
claimed a significant clustering of CR exposure ages measured for 80
iron meteorites, which they interpreted as periodic variations in the
CR background.

For potassium, the abundance ratio of two certain isotopes
($^{41}K/^{40}K$) is changed by the exposure to energetic CRs. A given
ratio thus determines the total CR exposure at a given level. The
total exposure time to CRs is the time meteoritic material after
breakup from a meteoroid parent body is exposed to CRs in its orbit
around the solar system, until it impacts on earth where it is
shielded from CR thereafter by the atmosphere. If an intrinsically
uniform time distribution of impacting meteorites were exposed to a
constant CR flux, one would measure a uniform distribution of exposure
ages.

If however the CR flux were variable the density of measured exposure
ages would appear modulated, and not uniform. Under the assumption of
a uniform intrinsic age distribution of meteorites, a non-uniform
measured distribution of CR exposure age means a variable mapping of
age to exposure age, and that the CR flux must have been variable in
the past (see S03 for a more detailed description).

S02, S03, and SV03 found a periodic clustering of measured CR exposure
ages in data of iron meteorites. They attempted to correct for the
effect of real, intrinsic age clustering for iron meteorites (see
Section~\ref{sec:cleaning} for more details) as the result of the
break-up of a meteoroid parent body into several meteorites
\citep[e.g., as discussed by][]{vosh67}.  From the resulting data they
claimed to find a significant clustering at a 143$\pm$10~Myrs period
with a probability that their periodic distribution was in reality
produced by a uniform distribution of only ``1.2\% in a random set of
realizations''. Using their original data source we repeat their
analysis, and discuss the following factors and their influence on the
results: The selection of the input data and use of different systems
of chemical groups in cluster cleaning (Section~\ref{sec:data}). The
implementation of the cluster cleaning algorithm
(Section~\ref{sec:cleaning}), and the influence of cleaning process on
the statistics (Section~\ref{sec:cleaninginfluence}). Finally the
influence of the exact size of the cleaning interval
(Section~\ref{sec:cleaninginterval}), of different age error models
(Section~\ref{sec:usingerrors}) and of number statistics
(Section~\ref{sec:numberstatistics}). We end this article with
discussion of the impact of these results and conclusions
(Sections~\ref{sec:discussion} and \ref{sec:conclusions}).

\section{The data}\label{sec:data}
The raw data base for exposure ages of iron meteorites used by S03 is
cited to be from \citet{vosh79} and \citet{vosh83}, while S02 used
only the former. Together these two publications provide data for 82
meteorites\footnote{S03 reports 80 meteorites, apparently disregarding
the oldest and one other object} with Fe age dating and ages in the
range $90\le t \le 2275$~Myrs. We list the combined data set in
Table~\ref{tab:data}, ordered by chemical group (see below) and
measured CR exposure age. For the two or three measurements,
respectively, of the Canyon Diablo, Norfolk, Rhine Villa, and Willow
Creek meteorites we compute a mean value, for Calico Rock we use the
newer value from \citet{vosh83}. One object is a Pallasite, a
stony-iron meteorite, also with Fe age dating. We include it in our
sample, as was done in S02. The cited age errors $\sigma(t)$ are taken
from the corresponding sources (see Section~\ref{sec:usingerrors} for
usage of different age error models). What we disregard here are newer
data from \citet{lavi99} who use different isotopes for dating 13
meteorites of this sample (but see Section~\ref{sec:usingerrors}). In
this article we however do not want to debate the database, but the
methods used by S02/S03.

The chemical groups listed in Table~\ref{tab:data} are the standard
chemical classifications for iron meteorites. `An' marks an anomalous
chemical composition that does not allow assignment to a standard group;
PAL is the one stony-iron meteorite. Since the time of publication the
classification scheme has been revised and currently 14 distinct
chemical groups for iron meteorites are recognised: IAB, IC, IIAB,
IIC, IID, IIE, IIF, IIG, IIIAB, IIICD, IIIE, IIIF, IVA, and IVB
\citep[see][and references therein]{wass02}. In this scheme the old
groups IA and IB are combined into IAB, IIA and IIB into IIAB, IIIA
and IIIB into IIIAB, and IIIC and IIID incorporated into IIICD. The
generally accepted interpretation is that different meteorites from a
group were part of the same parent meteoroid
\citep[e.g.,][]{vosh83,lavi99}, or could at least have formed from the
same input material. The 14 group scheme would leave members of 11
chemical groups in our sample, plus the anomalous irons, and one
stony-iron. In Table~\ref{tab:data} we give the classification in the
14 group scheme, as well as the original classification from
\citet{vosh79} and \citet{vosh83} in parentheses, where differing.

\section{The `100~Myr cleaning'}\label{sec:cleaning}
The break-up of a meteoroid into multiple meteorites, and their later
impact on earth, conflicts with the search for clusters in CR exposure
ages, since such groups represent real age cluster. The meteorites'
chemical composition can be used for an attempt to account for such
real age clusters, since meteorites from the same meteoroid parent
body should have a similar chemical composition.

S02/S03 suggested a correction for real age clustering using the
chemical classification. The specific methodology was likely motivated
by statements of \citet{vosh67} and \citet{vosh79}, who claimed that
errors in the age estimates from $^{41}$K--$^{40}$K isotope dating
method still allowed a discrimination between groups of meteorites
with ages of at least 100~Myrs apart, given some constraints on the
quality of measurement. S02/S03 subsequently reported to have
``removed all meteorites that have the same classification and are
separated by less than 100~Myr'' (S03) in age, and replaced them with
their average age. In this article we do not want to discuss whether
such a cleaning routine is sufficient or not, we leave this to others.

We identify two fundamental requirements to any such filter: (1) it
must be complete, i.e.,\ it must consider every data point exactly
once; and (2) it must work without any a priori assumption or input
with respect to the position of alledged clusters. While the
completeness in (1) is an obvious requirement, it is impossible to
reconcile it with a request for a uniqueness of the filter. Take the
example of the (fictual) age sequence of $t_1=200$, $t_2=250$, and
$t_3=310$~Myrs for three meteorites of a given chemical group. Both
$t_1$ and $t_3$ are within 100~Myrs of $t_2$, but not of each
other. To combine ages within 100~Myr of each other either $t_1$ and
$t_2$ can be averaged or $t_2$ and $t_3$. There is no preference for
either choice and with a maximum of 20 group members in the meteoritic
dataset such ambiguities are real and not only academic.

While S02 and S03 lacked a description, their procedure was
implemented as follows (N.~Shaviv, pers.\ comm.; `hierarchical
implementation'): For each chemical group the pair with smallest age
difference was determined and the ages averaged, weighted by their
errors, and new weight-errors computed from combining the two
errors. This was repeated for the next closest age pair, including
points from previous averaging steps, until no pairs with age
differences less than 100~Myrs are left.

This has the advantage to provide a receipe for the treatment of the
case above and to guarantee to find all singular pairs, but as a
consequence it will combine age points that had originally a larger
separation than 100~Myrs. As one example we could again use the
fictual three values $t_1$, $t_2$, $t_3$ we constructed above. When
assuming identical errors all three values would be combined into one.

We want to use requirement (2) to define a filter that fulfills
requirement (1) and replaces ages with less than 100~Myrs of each
other by the average, but does not create averages of averages, rather
only averages from original data points. This can be done by placing
100~Myr intervals on the time axis. In this case individual meteorites
inside such an interval can have partners outside the interval, less
than 100~Myrs apart, as demonstrated.

We see two possibilities for a 100~Myr interval placement without a
priori assumptions: First, consecutive 100~Myr intervals, without gap
and no assumptions made. This would impose a regular grid upon the
data, but could not guarantee that solitary pairs of ages less than
100~Myrs apart would be treated correctly. The second and adopted
possibility (`interval implementation') is a sequence of 100~Myr
intervals, each starting at the position of a data point: Starting
with the youngest object in a chemical group one would average all
objects within $<$100~Myrs of its age, then move to the next youngest
object outside this range and continue. In this way all solitary pairs
would be found. Indentically valid is a start at the oldest object and
interval placement towards younger ages.

All three schemes will modify the distribution statistics and could,
by aliasing effects of interval size and folding period, affect the
significance statistics, when folding over a supposed period. This has
to be taken into account when constructing statistical tests, we will
estimate its effects by simulations in the next sections.

When we apply the filter to the meteorite age dataset, using the
standard classification scheme of 14 groups described above, we
receive the resulting ages and errors given in
Table~\ref{tab:data}. We give new ages and age errors for both
versions of the interval implementation, from youngest to oldest age
($t_\mathrm{100,+}$, $\sigma(t_\mathrm{100,+})$), and oldest to
youngest ($t_\mathrm{100,-}$, $\sigma(t_\mathrm{100,-})$), as well as
the hierarchical implementation ($t_\mathrm{H}$,
$\sigma(t_\mathrm{H})$).

\begin{figure*}
\includegraphics[angle=-90,clip,width=\textwidth]{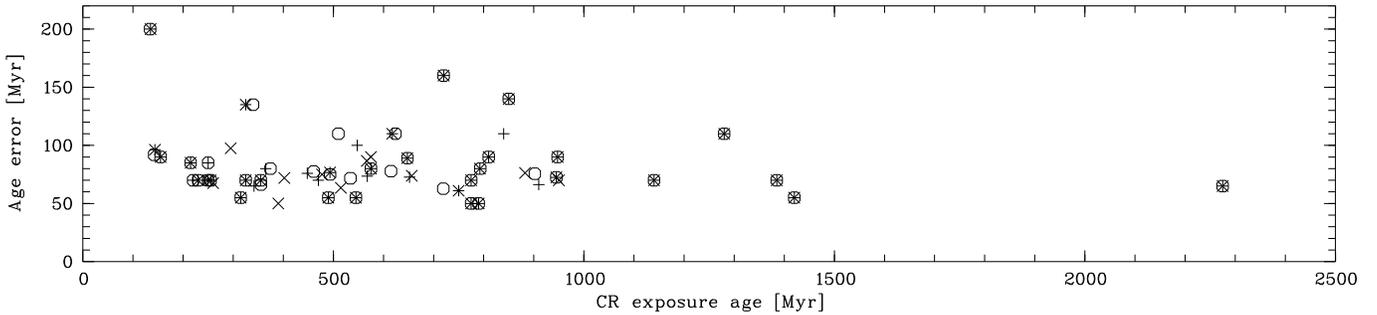}
\caption{\label{fig:datacleaned} 
Datasets of CR exposure age and age error after application of the
$<$100~Myr cleaning procedure. Results for the interval
implementation, filtering from youngest to oldest age ($\times$, 42
points) and from oldest to youngest ($+$, 43 points), and the
hierarchical implementation ($\circ$, 41 points)
}
\end{figure*}

After the $<$100~Myr cleaning with the above procedures, 42, 43, and
41 data points are left (of 82), respectively, compared to 50 of 80
for S03. The distributions of ages and errors derived with the
cleaning procedure described above are shown in
Figure~\ref{fig:datacleaned} for both filtering directions. They are
distinctly different from the points shown by S03. This is due to the
use of different chemical classification schemes. We use the current
modern classification while in S02/S03 the formal chemical
classifications were used as given in the original literature, without
combining related groups (N.~Shaviv, pers.\ comm.). Independent of
cleaning implementation, our data do not show strong apparent
clustering after cleaning. On the contrary, between 100~Myrs and
1000~Myrs the distribution appears rather uniform to the eye (see
Section~\ref{sec:vstatmeth}). However, we will quantify this statement
now.

\section{Exposure age statistics}\label{sec:statistics}
\subsection{Distribution tests}\label{sec:ks_kuiper}
The basic claim of S02/S03 was that a repeated clustering exists in the
CR exposure age data at 143$\pm$10~Myrs period. They reason that the
original dataset was cleaned to account for real age clustering and
subsequently folded over the proposed 143~Myr period. This folded
distribution was tested against a uniform distribution by a
Kolmogorov-Smirnov (KS) test. The KS test measures the maximum
distance $d$ between two cumulative distributions, and the KS
statistics converts this into a probability that the one distribution
has been randomly drawn from the test distribution
\citep[e.g.,][]{pres95}.

According to the test based on the KS statistics reported in
S02/S03, the folded data distribution was a chance realisation when
drawn from a uniform distribution with only 1.2\% probability
(identically for the two different datasets used in S02 and
S03). This 1.2\% probability was then interpreted as being a
significant sign of a deviation from a uniform distribution and, with
the folding step, that a periodic clustering of 143$\pm$10~Myrs was
present in the data.

We have to note here that the KS test is most sensitive to differences
around the mean of the distribution and less sensitive at the extreme
ends \citep{pres95}. In the case of periodic coordinates, as in the
present case when considering the folded data, the phase zeropoint can
be freely chosen to maximize the KS test signal. While S02/S03 did not
comment on the phase zeropoint used, it did not lie exactly at
$t=0$~Myrs and was likely shifted to achieve a maximum signal, which
is a valid step.

To avoid this arbitrary shifting, it is useful to employ a variant of
the KS statistics, the Kuiper statistics
\citep[e.g.,][]{pres95,step70}, that is an extension of the KS
approach to circular coordinates. It is independent of the phase
zeropoint of the independent coordinate, and thus more sensitive to a
signal at any position compared to the KS test. Instead of the maximum
distances $d$ between two cumulative distributions for the KS
statistics, the Kuiper statistic is based on the sum $v$ of the
maximum positive and negative distances between the two distributions.
In the following we will use the Kuiper statistics and its measure
$v$.

\subsection{Effect of the age cleaning}\label{sec:cleaninginfluence}
When the cleaning procedure is applied to remove signatures of real
age clustering (however physically appropriate), it does change
the distribution of the data points. In the extreme case, if the data
points were sampling the range of ages densely, the cleaning mechanism
would result in exactly one data point every 100~Myrs for the interval
implementation. For the hierarchical implementation the intervals
would be larger, with a size depending on the age errors. In the real
data the individual chemical groups only sparsely sample the age
interval.

To assess the magnitude of this influence we perform Monte-Carlo
simulations. We create random sets of 82 age points, uniformly drawn
from an age range of 0--1001~Myr (7 full periods for 143~Myrs).  We
then assign to each `object' age a random chemical group (out of the
11 present in the data), distributed as for the real data (1 to 20 per
group). Then the two implementations of the cleaning procedure are
applied as described above (only young-to-old for the interval
implementation) and the dataset folded over a 143~Myr period. We
repeat this 50\,000 times each. For each realisation the Kuiper test
against a uniform distribution is applied, to period-folded datasets
both with and without cleaning. While the $v$ measure itself depends
on the number of datapoints in a sample, the Kuiper probability
statistics do not and, thus, the two resulting distributions of
probabilities can be compared.

\begin{figure*}
\includegraphics[clip,width=\columnwidth]{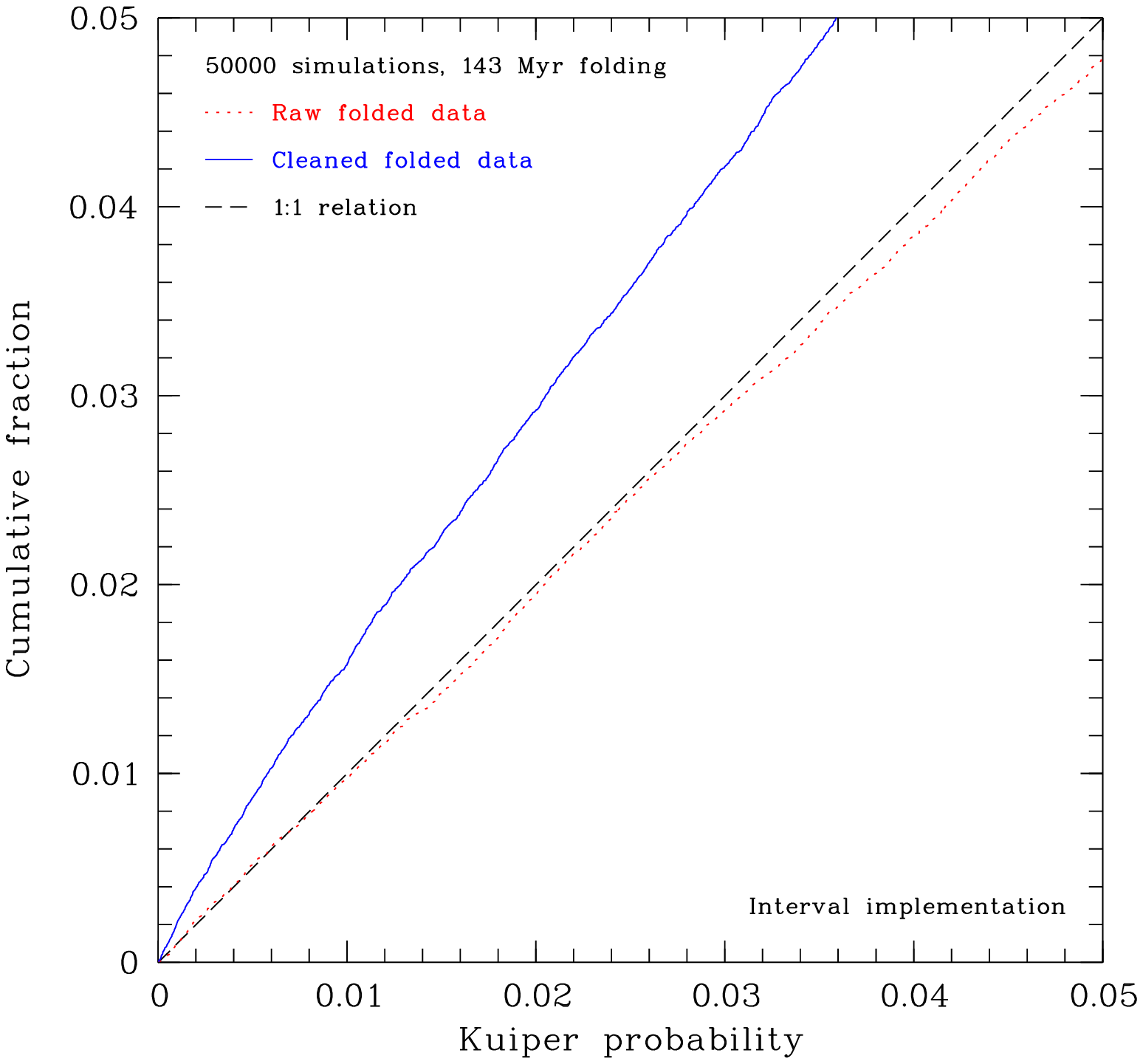}
\includegraphics[clip,width=\columnwidth]{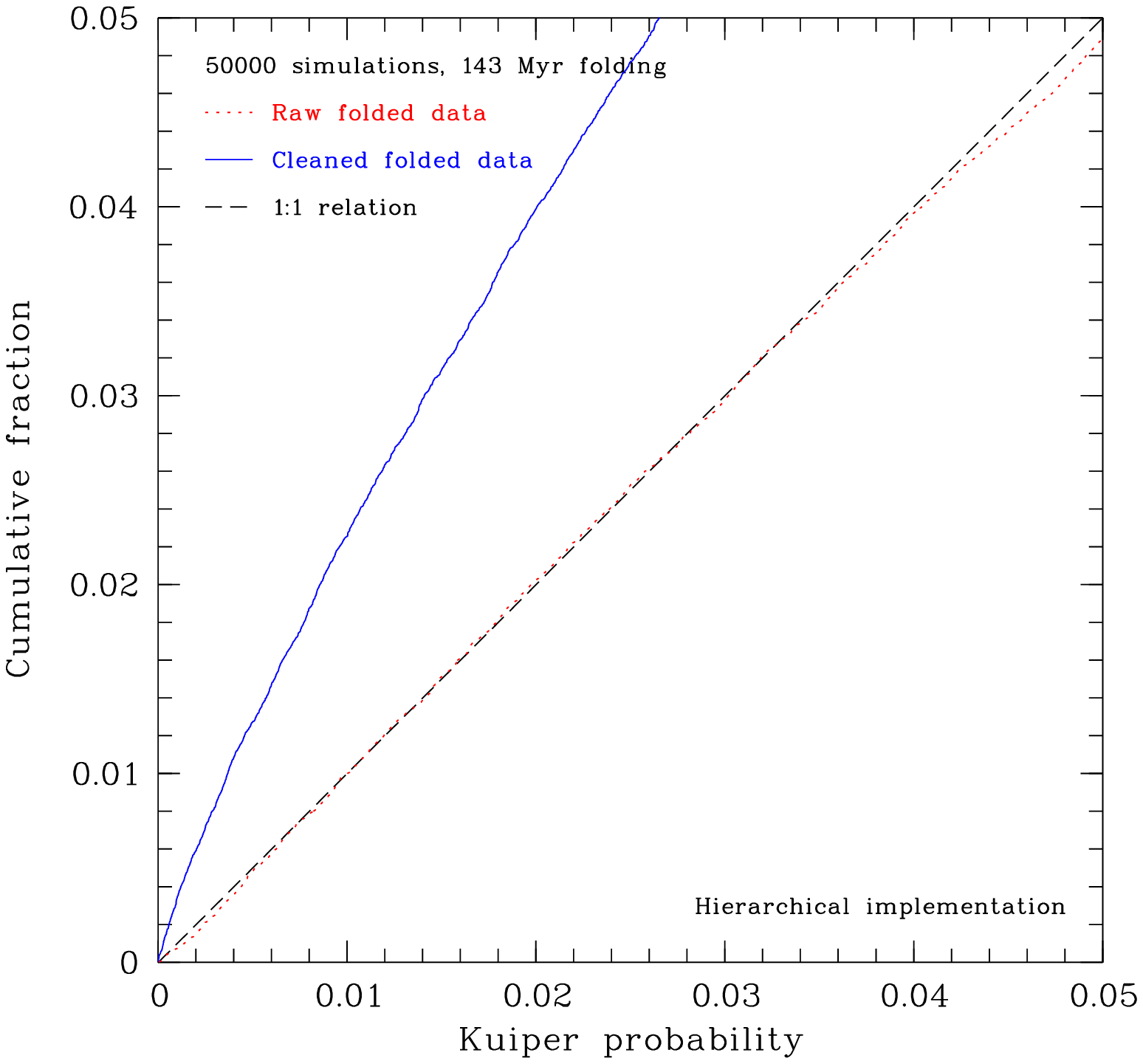}
\caption{
\label{fig:cleaning_bias}
Cumulative distributions of the Kuiper statistic probabilities for
50\,000 simulated datasets drawn from a uniform distribution, folded
over a 143~Myrs period, and compared with a uniform
distribution. Shown are the lowest, i.e.,\ most significant 5\%. Lines
mark the raw, uncleaned (dotted line) and cleaned distributions (solid
line). The dashed line marks the 1:1 limiting relation in the case of
infinitely many simulations and no influence of the filtering on the
distribution. Left: Interval implementation of the cleaning filter,
right: hierarchical implementation.
}
\end{figure*}

If the filter had no effect on the underlying distribution of
datapoints, the resulting probabilities should reflect the random
nature of the drawing process; i.e.,\ the derived distribution of
probabilities as given by the Kuiper statistics should again be
uniform. In Figure~\ref{fig:cleaning_bias} we plot the lowest 5\% part
of the cumulative distribution of the Kuiper test probabilities. 5\%
in this diagram means that in only 5\% of random realisations a
distribution as non-uniform as this should be drawn from a uniform
distribution.

For the raw, uncleaned distributions the probabilities lie as expected
close to the 1:1 relation. Deviations are due to statistical noise
and, to a small extend at the upper end, to the standard approximation
formula used in computing probabilities from $v$ \citep{step70}. This
is however not the case for the cleaned datasets, the probabilities
given by the Kuiper test are systematically too low -- for the
hierarchical implementation even lower than for the interval
implementation. This indicates that the distribution to compare
against after cleaning and folding is not anymore a uniform
distribution.

S02/S03 did not correct for this modified statistics, and if we assume
this simulation to be valid also for their different use of age
groups, their stated 1.2\% would have to be changed to
$\sim$3.7\%. This is not a very high significance level anymore. For
the interval implementation a 1.2\% value is measured for $\sim$1.8\%
of the cases.

What this initial test shows primarily is that the simple comparison
against a uniform distribution after cleaning and folding is not
valid. In order to test against a uniform {\em input} distribution,
the comparison distribution for a Kuiper test after cleaning and
folding would need to be somewhat similar to a uniform distribution
but its precise shape is not known. In the next section, we circumvent
this problem by continuing to compare to a uniform distributions to
compute the measure $v$, but we construct the distribution of
probabilities for the $v$ values from simulations, and we do not use the
Kuiper statistics directly.

\section{Statistical tests on the original data}\label{sec:datatest}

We want to study the probabilities that the two cleaned versions of
literature data as given in Table~\ref{tab:data} are drawn from a
uniform distribution, after folding over a given period. This not only
for a folding period of 143~Myr, but all periods ranging from 100 to
250~Myrs, to search for other periods with possibly significant
signals. While this will result in a single probability for each
period, we identified four factors that will result in an error bar on
these probabilities: (1) the two different cleaning implementation,
one with two cleaning directions; (2) the exact size of the cleaning
interval; (3) the age errors associated with the data; and (4) number
statistics.

\subsection{$v$-statistics for the real data}\label{sec:vstatmeth}

\begin{figure}
\includegraphics[clip,width=\columnwidth]{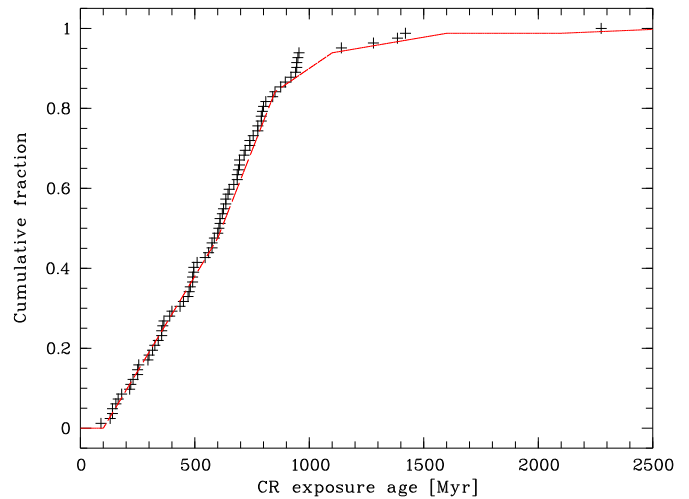}\hfill
\caption{
\label{fig:density}
Construction of artificial datasets. Cumulative distributions of
original data (symbols) and of function with $\ge$250~Myr piecewise
constant probabilities matched to the data (line). From this function
random datasets are drawn, with locally uniform distribution but this
general distribution function.
}
\end{figure}

For converting $v$ values into probabilities, we need to construct the
Kuiper $v$ statistic for datasets with similar properties as the
original data but drawn from a uniform distribution. While during
larger parts of the interval 100--1000~Myrs the distribution appears
to be rather uniform on larger scales, it clearly is not beyond
1000~Myrs (Figure~\ref{fig:datacleaned}). Therefore, as a basis for creating
artificial datasets we construct an age density distribution
that has a piecewise constant number density of meteorites, matched to
that of the real data. This is shown in Figure~\ref{fig:density}. The
intervals of constant number density have a size of $\ge$250~Myrs, which is
at least as large as the largest folding period that we test here. The
null hypothesis of the following tests is that the dataset is drawn
from this distribution, after cleaning and folding. We now draw
simulated datasets using a piecewise uniform distribution, with
probabilites proportional to the local number density. This is
identical to drawing random sets uniformly distributed in (0,1) and
translate these values to ages using Figure~\ref{fig:density}.

Datasets constructed in such a way have values locally distributed
uniformly, but follow the general density distribution on 250~Myr
scales; in this way no local clusters are created. We then assign
chemical classes to the 82 datapoints in each sample, with frequencies
as in the real data.

\subsection{Probabilities for cleaned data}\label{sec:realdata}

We construct 2500 artificial datasets as described above and cleaned over a
100~Myr interval, fold each dataset over periods of 100--250~Myrs in
1~Myr steps, and compute the $v$ values when comparing to a uniform
distribution. The same is done for the real data. This is repeated for
the three variations of the cleaning filter for both real data and
simulations. The comparison of $v$ for the real data with the $v$
statistics of the simulations determines the probabilities that the
former is only a random realisation of the latter.

\begin{figure*}
\includegraphics[clip,width=8.5cm]{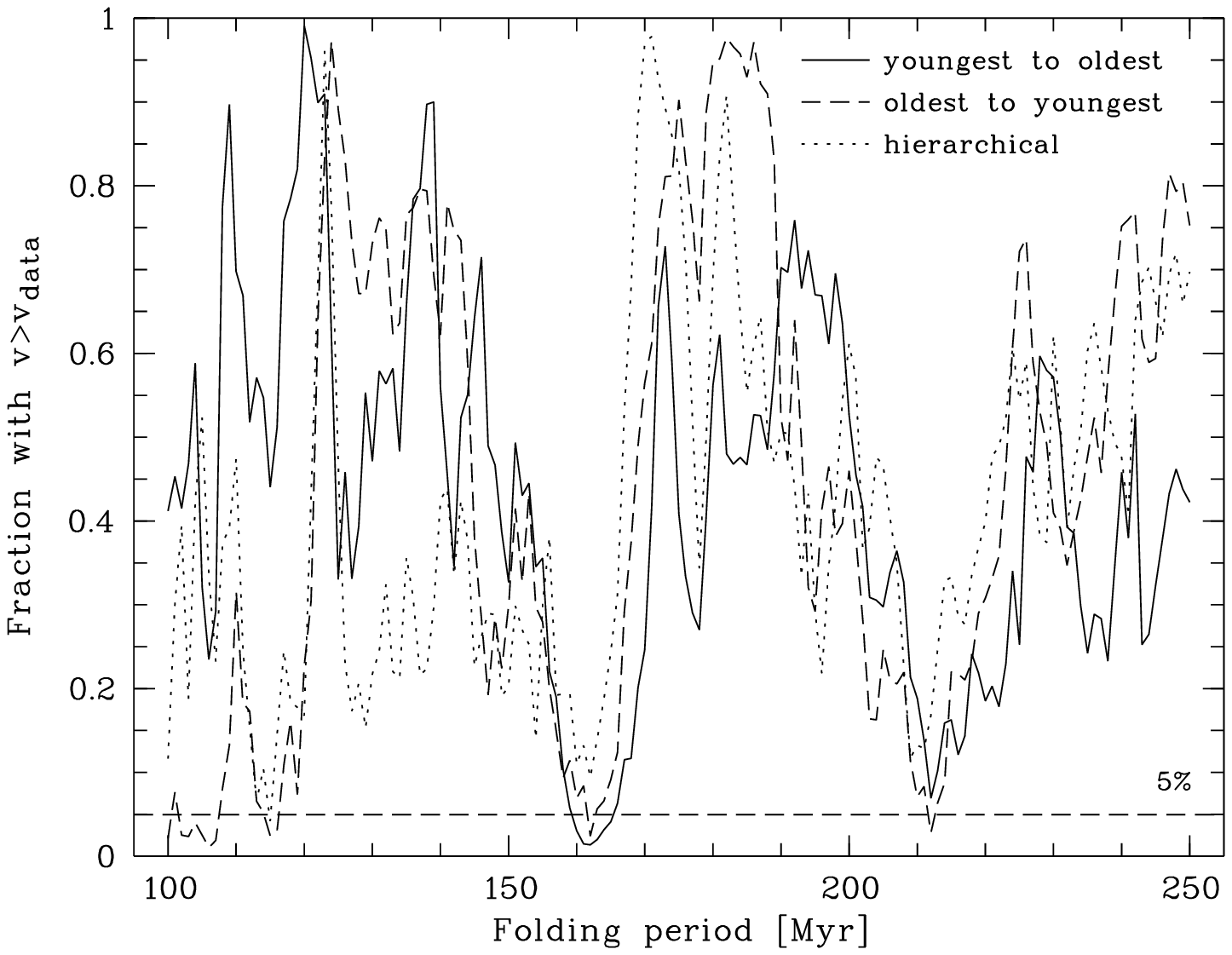}\hfill
\includegraphics[clip,width=8.5cm]{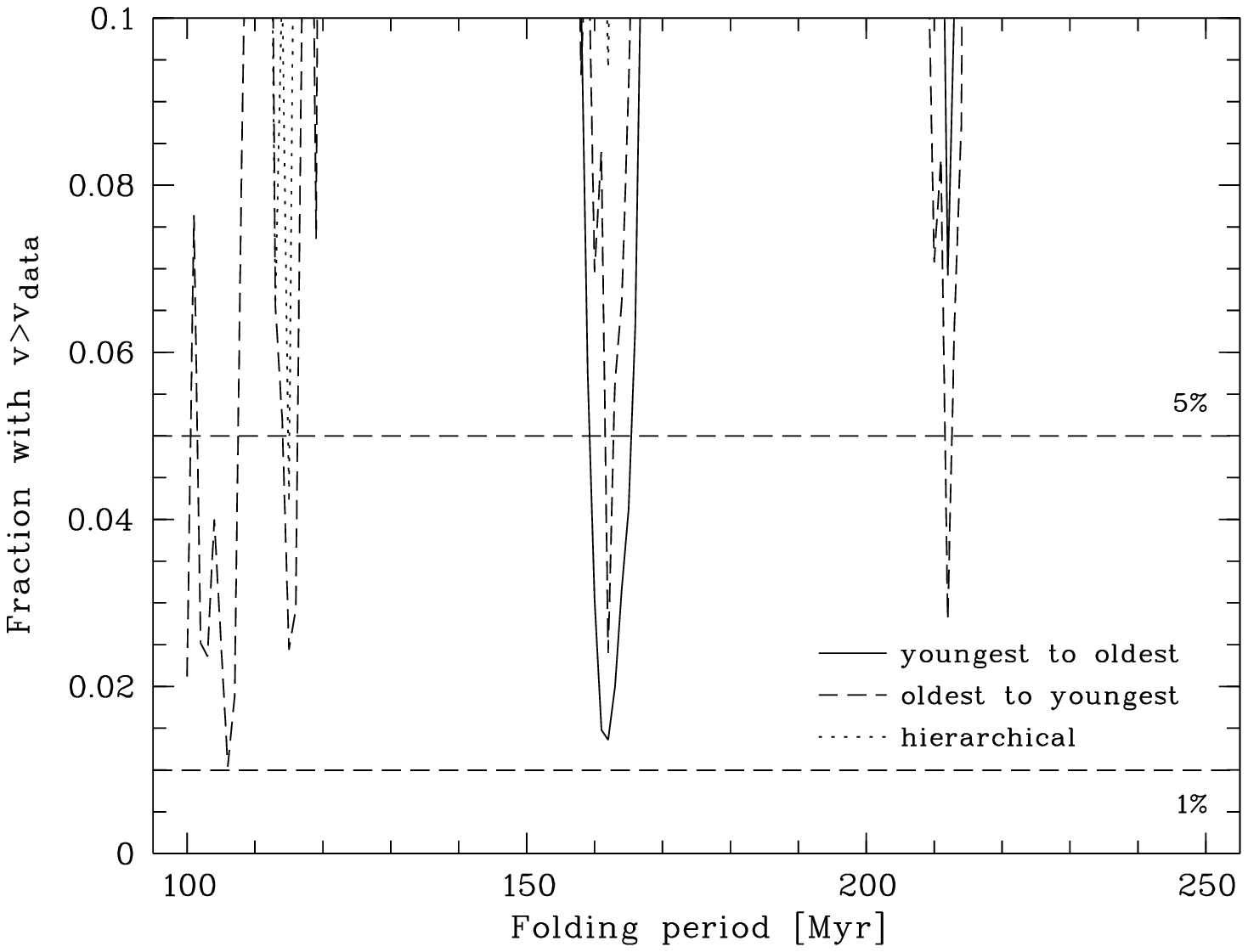}
\caption{
\label{fig:kuiper100_500}
Probabilities that the period-folded data are drawn from a uniform
distribution, as a function of folding period and without any error
bars. Shown are the full range of periods (100--250~Myrs) and full
range of probabilities (left) and the most significant 10\%
probabilities as a zoom (right). The three lines correspond to the two
interval cleaning variants, youngest to oldest (solid line) and oldest
to youngest age (dashed line), the dotted line to the hierarchical
cleaning (dotted line). Horizontal lines mark the 95 and 99\%
significance levels, respectively.
}
\end{figure*}

The resulting probabilities are shown in
Figure~\ref{fig:kuiper100_500}, the three lines indicate the three
cleaning variants. The relations deviate substantially for large parts
of the period, showing differences between a few and $>$50 percent
points. Over the full range the probabilities for all cleaning
directions reach below 10\% only around 162~Myrs.

\section{Sources of uncertainty}\label{sec:uncertainties}
While in Figure~\ref{fig:kuiper100_500} already the influence of the
different cleaning implementation is indicated, the next step is to
construct error bars on the probabilities reflecting also the other
three sources or error. We make the assumption that these are at
maximum weakly dependent on each other, and treat them separately.

\subsection{Size of the cleaning interval}\label{sec:cleaninginterval}
So far we used the cleaning interval size of 100~Myrs as suggested by
S02/S03. However 100 has a single significant figure -- the value is
not 100.0 -- which seems adequate since the value stems from a rough
estimate in \citet{vosh79}. For this reason we study the dependence of
probabilites on the exact interval size. We vary it by 10\%, so using
also 90 and 110~Myrs.

\begin{figure*}
\includegraphics[clip,width=8.5cm]{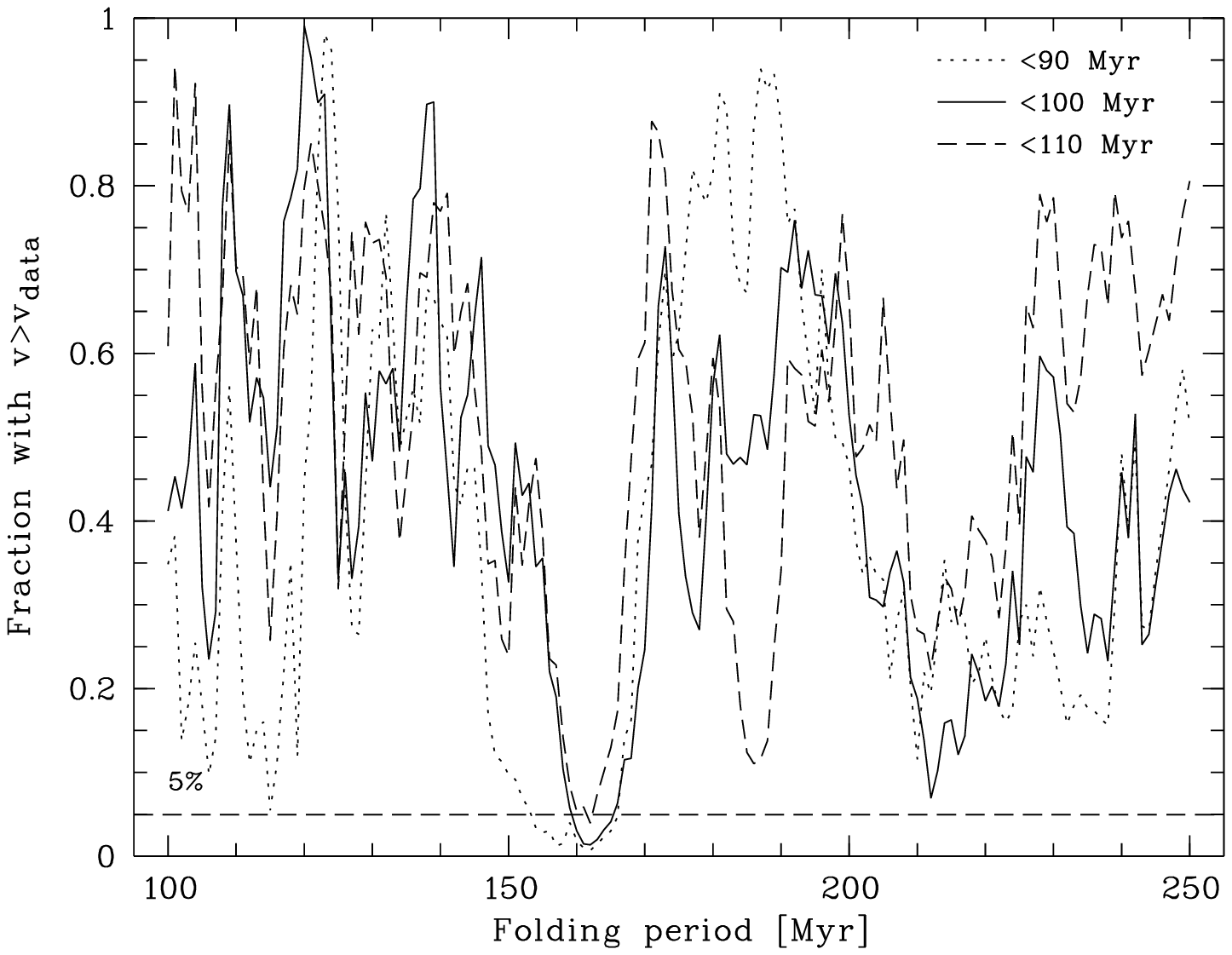}\hfill
\includegraphics[clip,width=8.5cm]{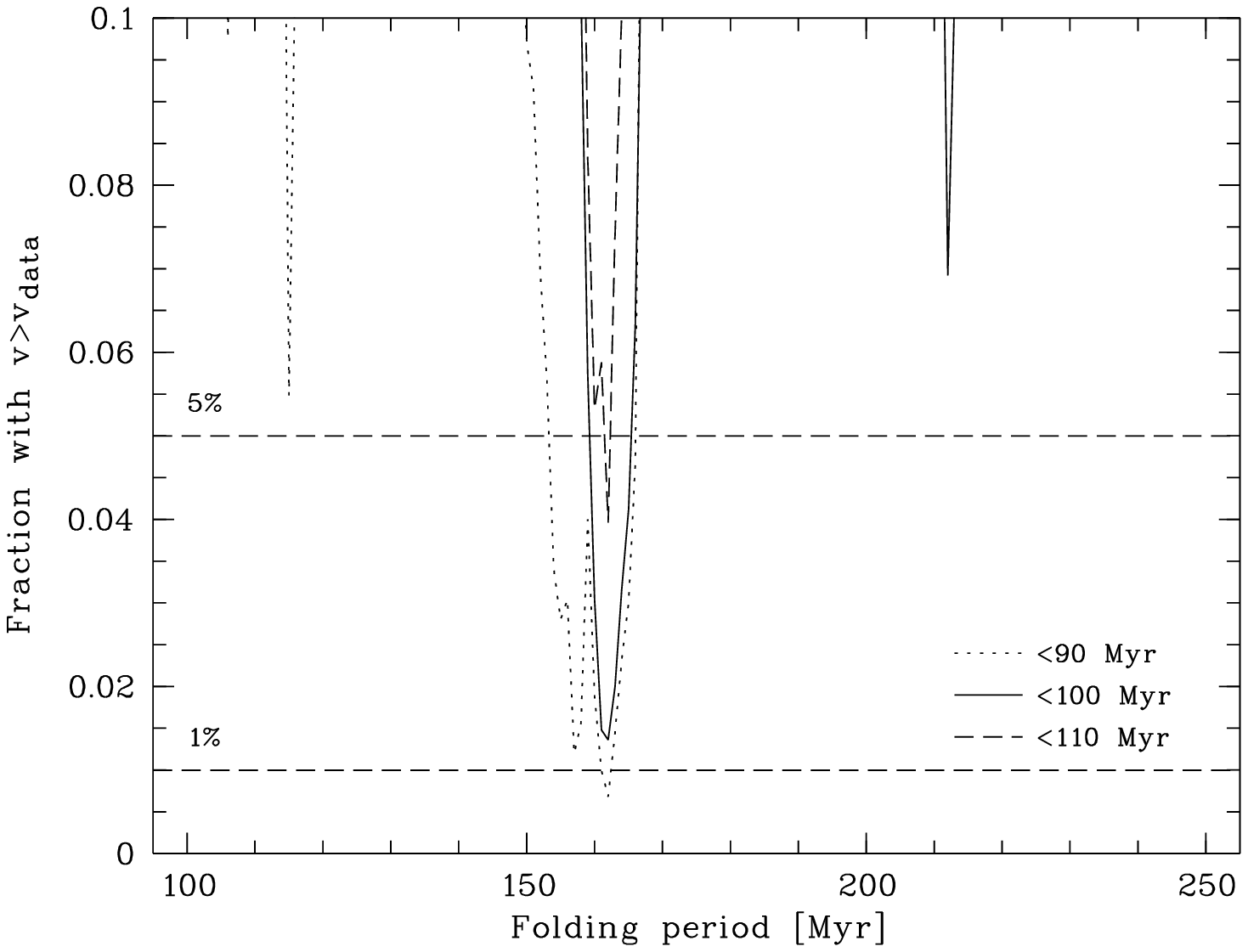}\\
\includegraphics[clip,width=8.5cm]{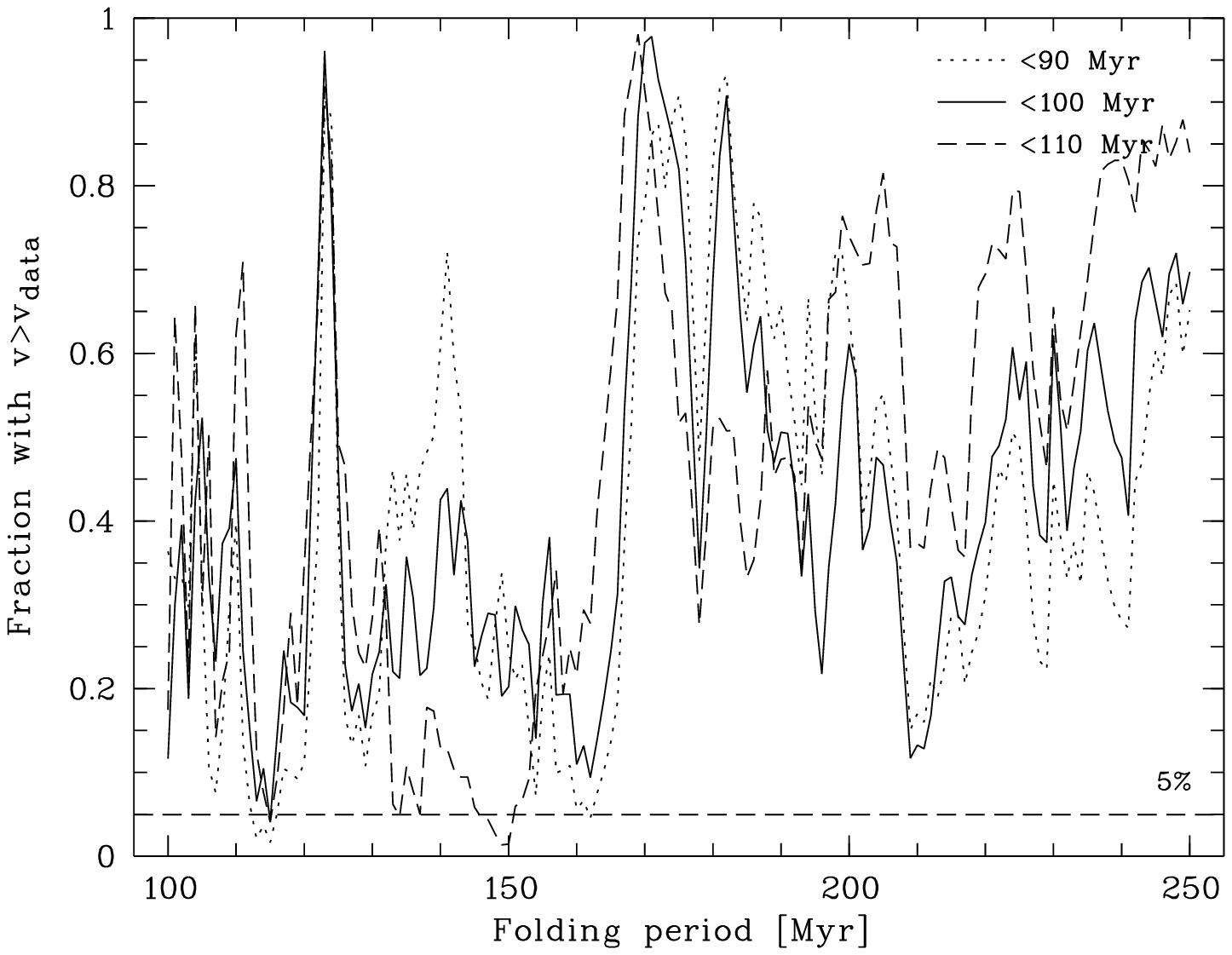}\hfill
\includegraphics[clip,width=8.5cm]{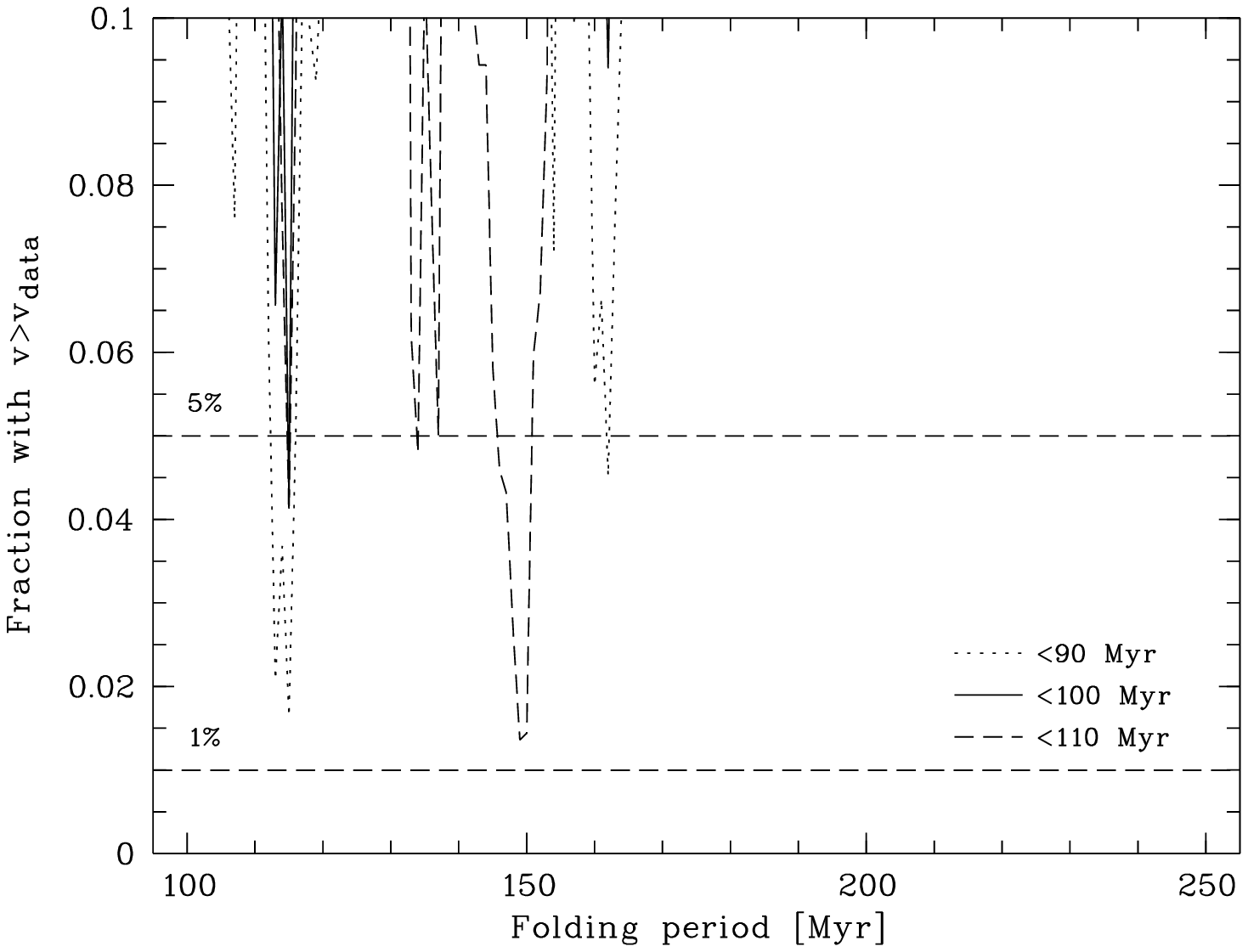}
\caption{
\label{fig:cleaninginterval}
Similar to Figure~\ref{fig:kuiper100_500} but for three cleaning
interval sizes: 90~Myrs (dotted line), 100~Myrs (solid line, as in
Figures~\ref{fig:kuiper100_500}), and 110~Myrs (dashed line). Top:
interval cleaning from youngest to oldest age; bottom: hierarchical
cleaning.
}
\end{figure*}

With these interval sizes we repeat the analysis from
Section~\ref{sec:realdata} above, again creating 2500 simulated
datasets and computing the $v$ statistics for simulations and real
data. We do this for the young to old interval cleaning and the
hierarchical cleaning. As shown in Figure~\ref{fig:cleaninginterval},
the results exhibit a spread between the three models of similar size
as for the use of different cleaning implementations.

\subsection{Age uncertainties}\label{sec:usingerrors}
The age uncertainties were neglected up to now. The errors in age
originally quoted by \citet{vosh79} and \citet{vosh83} lie in the
range $50<\sigma(t)<230$~Myr, and after the 100~Myr cleaning at
$50<\sigma(t_\mathrm{100})<200$~Myr. However, in a discussion of the
strength of the clustering signal, S03 claimed an ``at most 30~Myrs''
uncertainty in the ages, estimated from ``comparing the potassium ages
to ages determined using other methods''. While he did not give a
reference for this claim there, he likely refered to \citet{lavi99}.
That study showed substantially different CR exposure ages using
$^{36}$Cl, $^{36}$Ar, and $^{10}$Be measurements instead of $K$ for 13
meteorites. When following the conclusions in \citet{lavi99} of an
increase in CR flux over the last $\sim$10~Myrs, the ages from the two
isotope methods can be brought into better agreement and a comparison
delivers age uncertainties from comparing $K$ and $^{10}$Be ages of
rather 10--70~Myrs than 50--230~Myrs as in \citet{vosh79} and
\citet{vosh83}.

In S03 it was noted that ``the error will have the tendency to smear
the distribution''. This is an important point in the light that
S02/S03 disregarded the errors in the analysis altogether and did not
test their influence on the results.

\begin{figure*}
\includegraphics[clip,width=8.5cm]{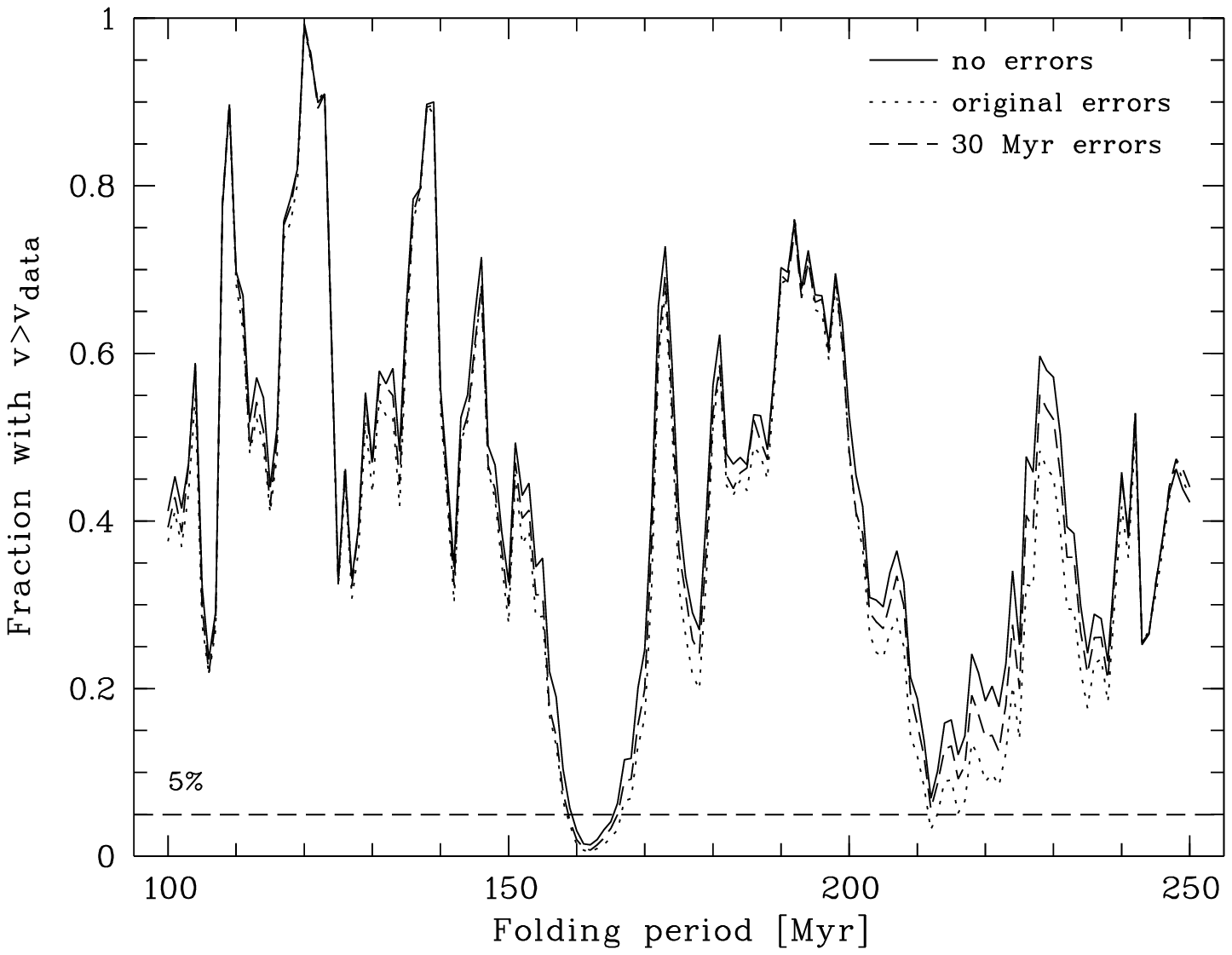}\hfill
\includegraphics[clip,width=8.5cm]{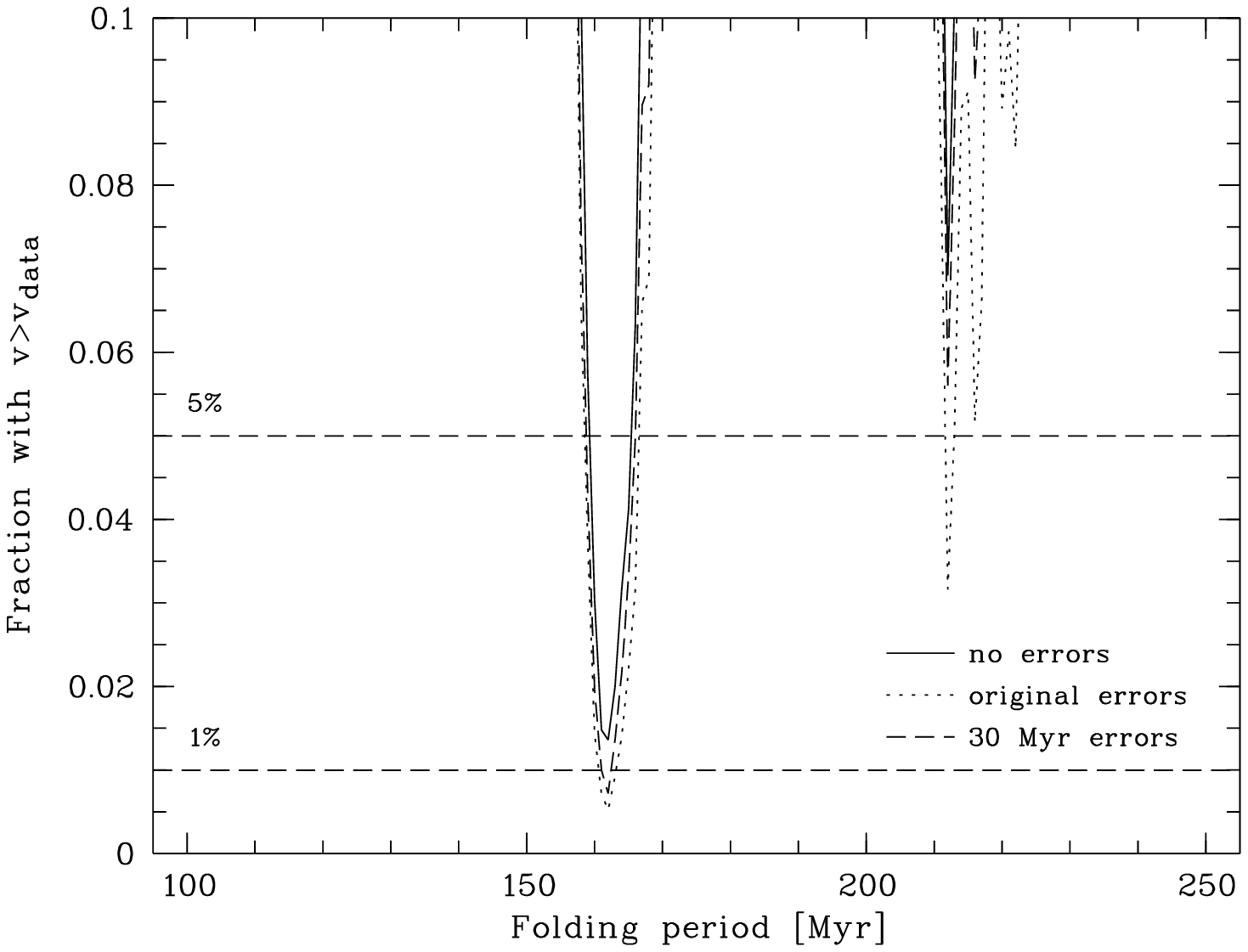}
\caption{
\label{fig:kuiper_errors}
As Figure~\ref{fig:kuiper100_500} but for different assumed age
uncertainty models. Data are interval cleaned (young to old,
100~Myr). Shown are the probabilities in case of no errors (solid
line, as in Figures~\ref{fig:kuiper100_500}), original errors (dotted
line), and assumed 30~Myr errors for all objects (dashed line).
}
\end{figure*}

While we again can not and do not attempt to determine `real' age
uncertainties, we want to assess the contribution to probability error
bars from different models of age uncertainties. We again use
simulation to create a statistic of $v$ values. Here we assume two
error models in addition to the `no errors' in
Section~\ref{sec:realdata}: (a) the originally published errors, and
(b) 30~Myr errors for all data points. We again create 2500 datasets
each, assuming the errors to be gaussian, age-clean (100~Myr), and
fold them to derive the $v$ statistics for these sets. The results are
shown in Figure~\ref{fig:kuiper_errors}, similar to
Figure~\ref{fig:kuiper100_500}. The chance realisation for a null
hypothesis increases by a small amount for the original errors
compared to the 30~Myr errors or the case without errors. In
comparison to other sources of uncertainty, the effect of age errors
is in fact negligible.

\subsection{Number statistics}\label{sec:numberstatistics}
The last source of error we study here is the influence of number
statistics on random clustering in the real data. With $\sim$40
datapoints in the dataset after cleaning, each individual point has a
non-negligible influence on the statistics.

So far, the $v$ distribution from simulated datasets shows the effect
of different discrete random realisations of the {\em null
hypothesis}. However, the simulation do not make statements about the
influence of number statistics from the {\em data side}. We need to
quantify how strongly an apparently significant deviation from the
null hypothesis might be depending on a single or a few data points,
i.e.,\ statistical outliers.

For this application the statistical method of bootstrap simulations
has been shown to be a valid approach \citep[see][and references
therein]{pres95}, given that the data are independent and identically
distributed. Even though this is not strictly the case here after the
application of the cleaning filter, the interdependences of datapoints
are both rather local and weak. We thus assume that this has only a
negligible influence, which allows us to perform a bootstrap of the
data.

The bootstrap allows the estimation of error bars for a certain
parameter from a measured dataset itself. New datasets with the same
size as the original are drawn from the original dataset with
replacements.  The parameter in question is then determined from the
bootstrapped datasets as before and the spread in this parameter is a
good estimate for its uncertainty.

Here we use bootstrapping to estimate the influence of number
statistics on the $v$ value for our data. $v$ is a valid parameter
with which to apply bootstrapping, but with the follwoing caveat: the
$v$ statistics gets skewed by the bootstrapping process itself, as a
result of some datapoints being present more than once in the
bootstrapped datasets. This changes the cumulative distributions to be
less smooth, and thus skewes the $v$ statistics towards higher values.

\begin{figure*}
\includegraphics[clip,width=8.5cm]{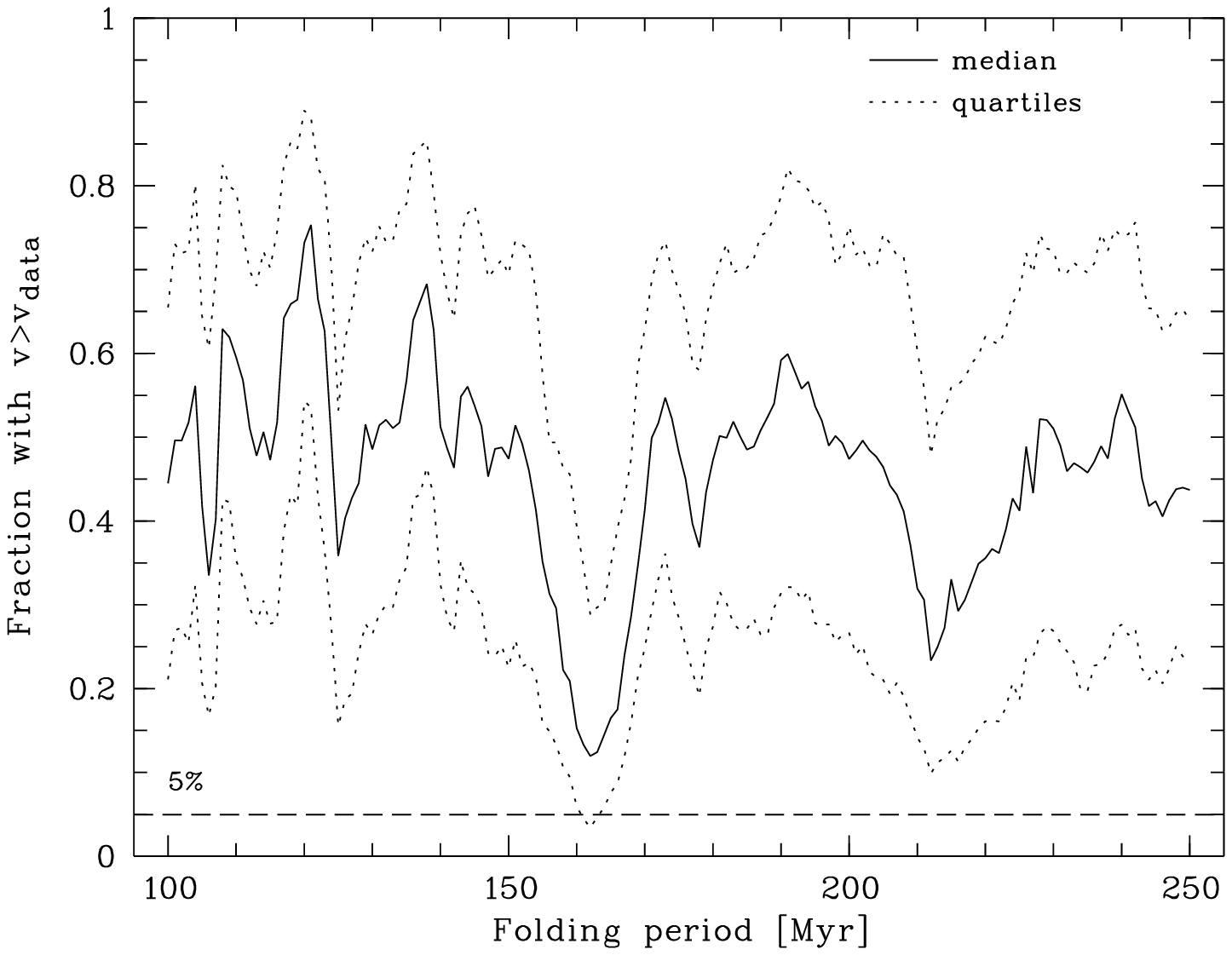}\hfill
\includegraphics[clip,width=8.5cm]{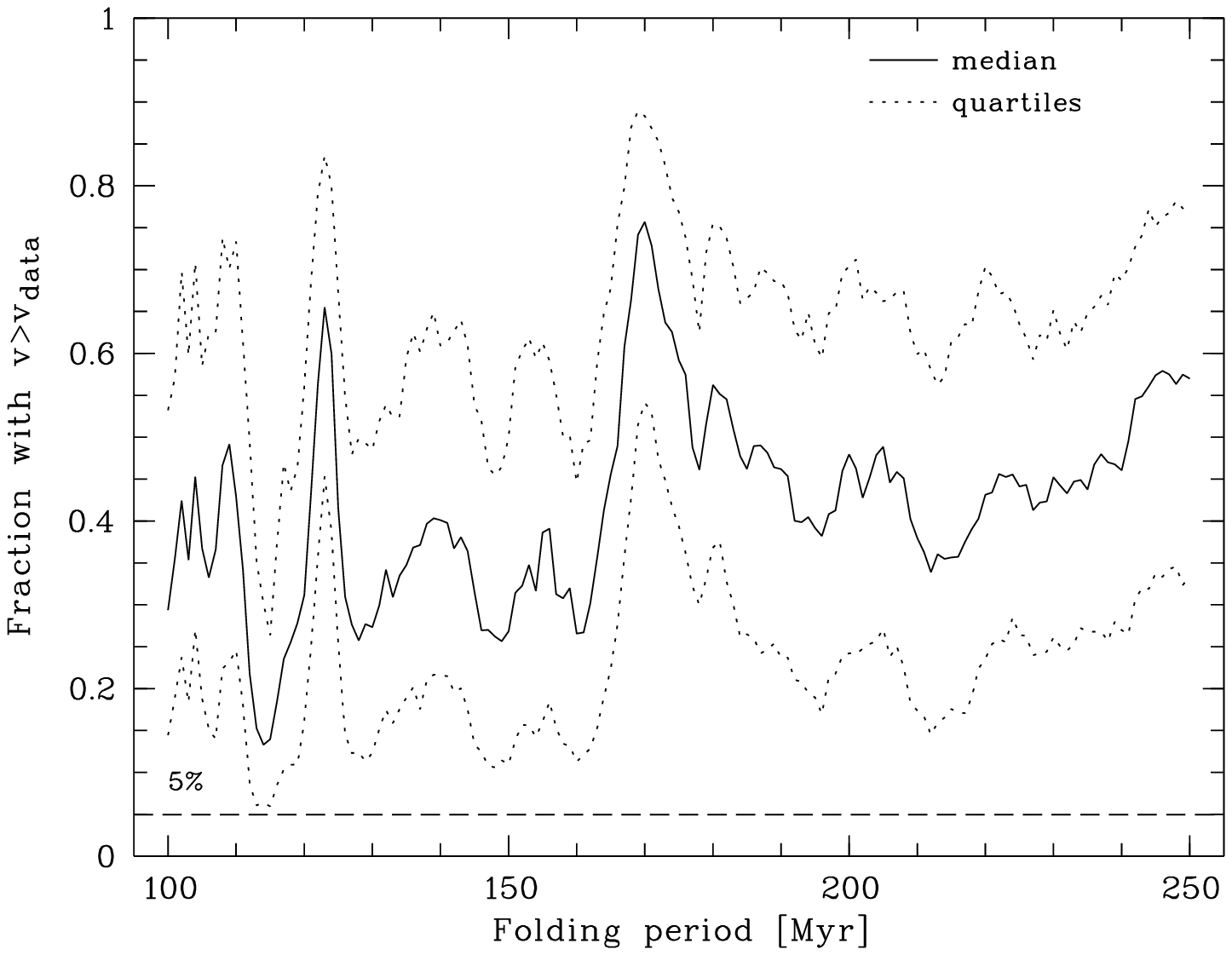}
\caption{
\label{fig:bootstrap_data}
Distribution of probabilities from bootstrapping of the 100~Myr
interval cleaned dataset, cleaned younger to older ages (left) and
hierarchical cleaning (right). Shown are the median (solid line) and
upper and lower quartiles (dotted lines) of the probabilities. As a
result of number statistics no period has a significant signal for
deviation from a uniform distribution.
}
\end{figure*}

For this reason we construct bootstrapped datasets from our data and
compare these to {\em bootstrapped} datasets of simulations. In this
way the same modification is applied to both sides, and $v$ can again
be compared. We create 2500 bootstrapped simulation for this case,
applying the bootstrapping after cleaning. The dataset is also cleaned
with a 100~Myr interval (interval implementation, younger to older
ages, and hierarchical implementation), and then bootstrapped 500
times. We fold the distributions over 100--250~Myr periods, and
receive a $v$ statistic for each period given the null hypothesis. By
comparing the $v$ statistics from the bootstrap realisations of the real
data to that of the simulations we get a distribution of probabilities
that the measured $v$ is larger than the random simulated $v$. We show
the median and upper and lower quartiles in
Figure~\ref{fig:bootstrap_data}.

The lower quartile drops to $\sim$5\% around 162~Myrs for the interval
implementation, and for the hierarchical implementation around
114~Myrs. For no period in both implementaions the median drops below
the 10\% mark. This indicates that in the tests above only a few
outlying datapoints are responsible for the probabilities below 5\%.

\section{Discussion}\label{sec:discussion}
Simple statistical tests against a uniform distribution do not tell
the whole truth about the clustering properties of the dataset of
meteorite CR exposure ages. It is obvious from
Figure~\ref{fig:kuiper100_500} that the exact implementation of the
cleaning mechanism can already have a strong influence on the
composition of the dataset.

A similar case is the influence of the exact choice of the cleaning
interval (Figure~\ref{fig:cleaninginterval}). There is no good
argument available from S02/S03 or the referenced literature why more
than one significant figure for the `100'-Myr interval should be
assumed. Thus 90 and 110~Myrs are valid variations -- 80 and 120~Myrs
would be, too -- with the substantial influence on the resulting
probabilities seen above. In comparison, the influence of the age
uncertainties using different age uncertainty models
(Figure~\ref{fig:kuiper_errors}) is small compared to the first two
sources.

\begin{figure*}
\includegraphics[clip,width=8.5cm]{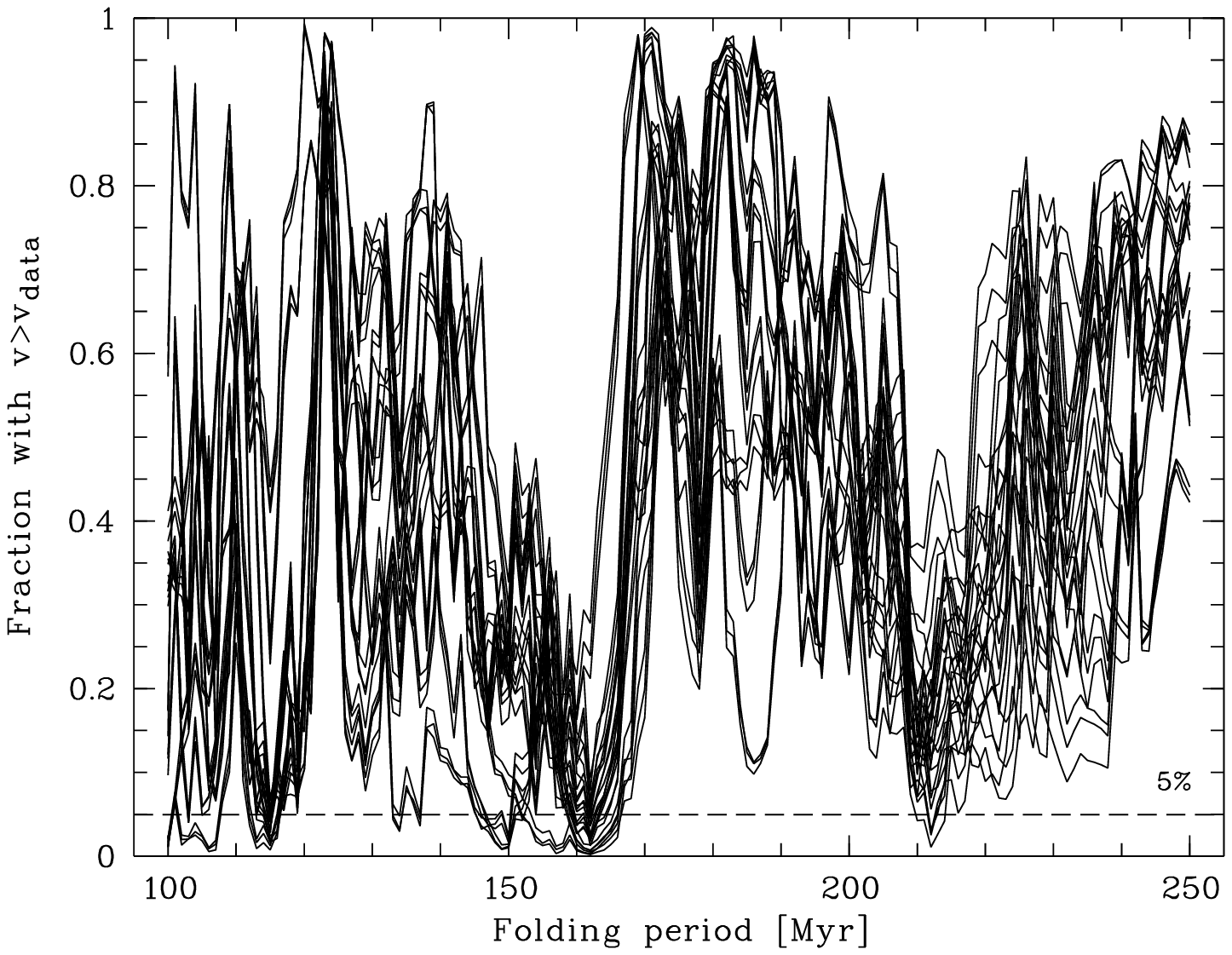}\hfill
\includegraphics[clip,width=8.5cm]{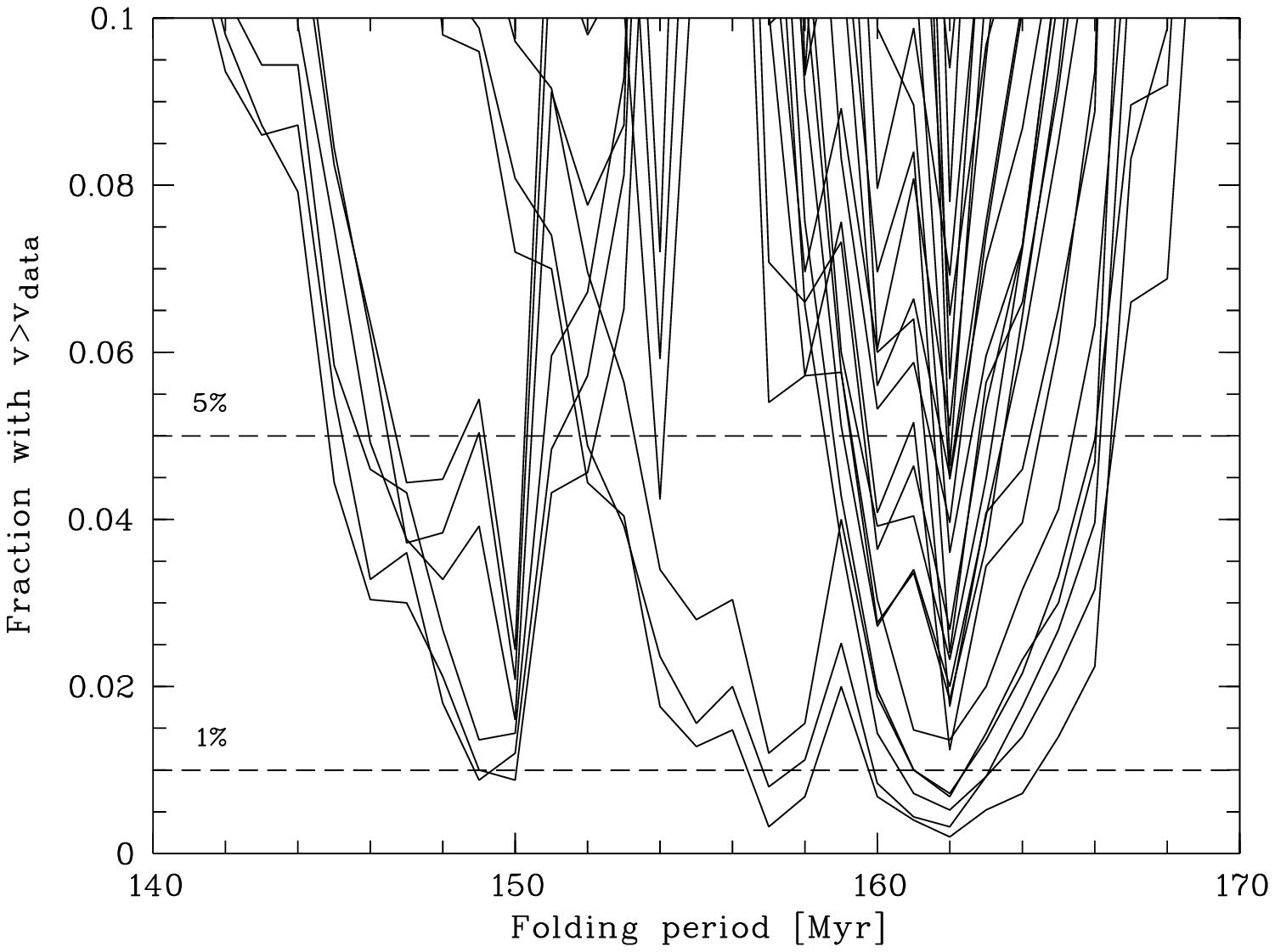}
\caption{
\label{fig:allprobs}
As Figure~\ref{fig:kuiper100_500} but for the 27 different
combinations of cleaning implementation, cleaning interval size, and
error model in Table~\ref{tab:162myr}. Left: Full range of periods
from 100 to 20~Myrs. Right: Zoom on region around 143 and 162~Myr,
showing the top 10\% probabilities.
}
\end{figure*}

In Figure~\ref{fig:allprobs} we show the probabilities for a total of
27 different combinations of cleaning implementation, cleaning
interval size, and error model, expored by simulations. The used
combinations are listed in Table~\ref{tab:162myr}. The existence of
such variations is a consequence of the fact that an inherent
clustering of real meteorite ages as a difference to CR exposure ages
has to be removed from the input data to allow a conversion of
exposure age distribution to CR flux levels. A filtering procedure for
this fact is not uniquely defined.

If the variations we make are valid, and the arguments given above
suggest so, then the spread seen in the probabilities for a given
period is a good indicator how significant the signal for a given
folding period is at maximum -- neglecting the influence of number
statistics for now. From Figure~\ref{fig:allprobs} it is clear that
there exists no significant signal for a deviation from a uniform
distribution of ages for a period of $\sim$143~Myrs. In only three of
the 27 cases the probabilities for a random realisation drops below
10\%, in none below 5\%. At a period of 150~Myrs there are six
combinations that give probabilities of around 1\%. However, the
remaining 21 are at $>$10\% and thus not significant. The only periods
that give low values lie around 162~Myrs.

We summarized the probabilities for the 162~Myr period in
Table~\ref{tab:162myr}. While for this period there are several
parameter combinations that result in probabilities below 1\%,
there are also others that are above 2.5, 5, and even 10\%. If there
are no strong arguments against these combinations as being valid,
then also the 162~Myr period clearly disqualifies as showing a
significant signal for being different from a uniform distribution. At
other periods there are also some of the parameter combinations with
probabilities below 5\% that are countered by combinations with above
10\% probability.

\begin{table*}
\caption{\label{tab:162myr}
Probabilities for a deviation from a uniform distribution when folding
at 162~Myrs, while {\em neglecting the influence of number
statistics}. Given are the probabilities for different cleaning
implementations (interval young to old, old to young ages, and
hierarchical), exact sizes of the cleaning interval (90, 100,
110~Myrs), and the assumed error model for the ages (no errors, 30~Myr
errors, original errors).
}
\centering
\begin{tabular}{llllllllll}
\hline\hline
Cleaning&\multicolumn{3}{c}{Interval: young to old}&
\multicolumn{3}{c}{Interval: old to young}&
\multicolumn{3}{c}{Hierarchical}\\
Error model&90~Myr&100~Myr&110~Myr&90~Myr&100~Myr&110~Myr&90~Myr&100~Myr&110~Myr\\
\hline
no errors	&0.68\%	&1.4\%&4.0\%	&6.4\%	&2.4\%	&7.8\%&4.5\%&9.4\%&27.8\%\\
30~Myrs		&0.32\%	&0.72\%&2.3\%	&4.5\%	&1.8\%	&5.7\%&2.7\%&6.9\%&23.9\%\\
original errors	&0.20\%	&0.52\%&2.0\%	&3.6\%	&1.2\%	&4.6\%&1.8\%&5.1\%&22.1\%\\
\hline
\end{tabular}
\end{table*}

Into this interpretation one additional factor enters: number
statistics. Figure~\ref{fig:bootstrap_data} demonstrates the span of
probabilities that is induced by number statistics, as tested by
bootstrap simulations. For the simulations we used both the 100~Myr
interval and hierarchical cleaning, and absence of errors. For a
different choice of the other parameters we expect some shifts in this
distribution, but no dramatic changes. The spread in probabilities in
Figure~\ref{fig:bootstrap_data} (shown are median and quartiles) is a
direct expression of small number statistics in the data. For the
shown cases we have 42 and 41 data points in the sample after
cleaning. With increasing number of data points this spread should
decrease. So in order to decrease this to a range that allows
detection of only 1\% probability, at the given strength of potential
signals, the dataset has to be larger by, say, at least an order of
magnitude.

In conclusion, we see no signal of periodic clustering with any period
between 100 and 250~Myrs for the dataset of 82 iron (including one
stony-iron) meteorites.  For all periods the dataset of CR exposure
ages is consistent with being drawn from a uniform distribution of
ages after cleaning for multiple-breakup clusters. This conclusion
holds when including all discussed error sources, and even when
incorrectly neglecting the effects of number statistics.

Why are these results differing so strongly from S02/S03? We identify
three main resons: 
\begin{itemize}
\item The implementation of the cluster cleaning
filter is clearly different between S02/S03 and this study, by using a
different chemical grouping scheme. However, there is agreement in
meteoritics on the current 14 group classification (plus possible further
extentions). In any case this allows meteorites (e.g.,\ from the former
IIIA and IIIB groups) to originate from the same parent body in the same
break-up event, which has to be recognised in the cleaning process.
This leaves us with less chemical groups and hence less data points
after cleaning, compared to S02/S03.

\item S02/S03 did not test the influence of the cleaning process on the
statistical properties of their dataset. This led to a skewed
statistic and falsly too low probability values even for their
original method.

\item In S02/S03 no check of the influence of error sources on the
face-value results of the KS-test was done. In particular, they did not
test the influence of the relative importance -- from small number
statistics -- of individual datapoints on their results. This together
resulted in an substantial over-interpretation of their result as
being significant.
\end{itemize}

These statements are made from a statistical side. We want to make
clear that there are other issues that we did not touch, e.g.,\
whether the proposed filter against intrinsic meteorite breakup
clustering is sufficient and thus useful. Residuals of intrinsic
meteorite clustering would of course strongly influence the detection
of CR exposure age clustering. Especially if the sought periodicity of
143~Myrs is only a factor of $\sim$1.4 longer than the proposed real
clustering length.

\section{Conclusions}\label{sec:conclusions}
We have investigated the claim by S02/S03 that a sample of $\sim$80
iron meteorites showed a CR exposure age distribution with a
143$\pm$10~Myr periodic clustering over the last 1--2~Gyrs. From this
they concluded a periodicity in the CR influx from different amounts
of star formation during the solar system's passage through the spiral
arms of our galaxy.

We followed their approach and computed the probability that the data
are drawn from a uniform distribution of ages, when folded over the
proposed period. As a difference to S02/S03 we studied several sources
that create uncertainties in the derived probabilities, and tested the
influence of filtering of their data, by using simulations.

The data are `cleaned' from real age clusters from breakups of
meteoroids into multiple pieces -- as suggested by S02/S03. As a side
result we find that such a filter can be implemented in several ways,
with all implementations having special advantages and
disadvantages. Computing the probabilities of the data as random
realisations of a uniform distribution we see a minimum at a period of
162~Myrs, and clearly not at 143~Myrs. When assessing the influence of
different sources of uncertainty, we compare the probabilities for a
random realisation for this 162~Myr period. When neglecting the
influence of number statistics to study the effects of the different
error sources, we find a non-negligible influence of (i) the
implementation of the age filtering, (ii) the exact choice of the size
of the cleaning interval, and (iii) to a smaller amount the influence
of different assumed age error models.

However, this is with the neglection of noise from number statistics.
There is no folding period with a consistent probability for a random
realisation of a uniform distribution of below 5\%, when considering
the above error sources, including 143 and 162~Myrs period.

On top of this, number statistics is clearly the strongest source of
influence, larger than the three sources above. Noise from small
number statistics -- $\sim$40 data points in the sample after cleaning
-- creates a scatter in the probability of the data, being a random
realisation of an underlying distribution. For any folding period from
100 to 250~Myrs $\ga$75\% of the corresponding bootstrap realisations
created for the dataset deliver probabilities for a random draw from a
uniform distribution of 5\% or higher, including the 143 and 162~Myr
periods. Thus, there is no period between 100 and 250~Myrs at which
the folded age distribution of the dataset is inconsistent with being
drawn from a uniform distribution. With the data and the methods
proposed by S02/S03 no periodic variation of the cosmic CR background
is found.

The differences of interpretation in S02/S03 to our results are due
to: (i) the use of an outdated chemical classification scheme, (ii)
the neglection of the influence of the filtering against real age
clusters on the KS statistics, and (iii) the neglection of error
sources, including number statistics, on the significance of the
results.

\begin{acknowledgements}
I would like to thank Lutz Wisotzki, Bj\"orn Menze, and Dan~H.\
McIntosh for fruitful discussions and suggestions. A special thanks
goes to Henning L\"auter for his critical review of my bootstrap
approach. I am grateful to Nir Shaviv for providing background
information on his methodology.
\end{acknowledgements}

\bibliography{knuds}

\begin{thebibliography}{17}
\expandafter\ifx\csname natexlab\endcsname\relax\def\natexlab#1{#1}\fi

\bibitem[{de~la Fuente~Marcos \& de~la Fuente~Marcos(2004)}]{fuen04}
de~la Fuente~Marcos, R. \& de~la Fuente~Marcos, C. 2004, New Astronomy, 10, 53

\bibitem[{{Gies} \& {Helsel}(2005)}]{gies05}
{Gies}, D.~R. \& {Helsel}, J.~W. 2005, ApJ, in press, astro-ph/0503306

\bibitem[{{Kristj{\' a}nsson} {et~al.}(2002){Kristj{\' a}nsson}, {Staple},
  {Kristiansen}, \& {Kaas}}]{kris02}
{Kristj{\' a}nsson}, J.~E., {Staple}, A., {Kristiansen}, J., \& {Kaas}, E.
  2002, Geophysical Research Letters, 29, 22

\bibitem[{{Laut}(2003)}]{laut03}
{Laut}, P. 2003, Journal of Atmospheric and Solar-Terrestrial Physics, 65, 801

\bibitem[{Lavielle {et~al.}(1999)Lavielle, Marti, Jeannot, Nishiizumi, \&
  Caffee}]{lavi99}
Lavielle, B., Marti, K., Jeannot, J., Nishiizumi, K., \& Caffee, M. 1999, Earth
  and Planetary Science Letters, 170, 93

\bibitem[{Press {et~al.}(1995)Press, Teukolsky, Vetterling, \&
  Flannery}]{pres95}
Press, W.~H., Teukolsky, S.~A., Vetterling, W.~T., \& Flannery, B.~P. 1995,
  Numerical recipes in C, 2nd edn. (Cambridge University Press)

\bibitem[{Rahmstorf {et~al.}(2004)Rahmstorf, Archer, Ebel, Eugster, Jouzel,
  Maraun, Neu, Schmidt, J., Weaver, \& Zachos}]{rahm04}
Rahmstorf, S., Archer, D., Ebel, D.~S., {et~al.} 2004, Eos (Transactions,
  American Geophysical Union), 85, 38

\bibitem[{Shaviv(2002)}]{shav02}
Shaviv, N. 2002, Phys. Rev. Letters, 89, 051102

\bibitem[{Shaviv(2003)}]{shav03}
---. 2003, New Astronomy, 8, 39

\bibitem[{Shaviv \& Veizer(2003)}]{shav03b}
Shaviv, N. \& Veizer, J. 2003, GSA Today, 13, 4

\bibitem[{Stephens(1970)}]{step70}
Stephens, M.~A. 1970, Journal of the Royal Statistical Society, Series B, 32,
  115

\bibitem[{Svensmark(1998)}]{sven98}
Svensmark, H. 1998, Phys. Rev. Letters, 81, 5027

\bibitem[{Voshage(1967)}]{vosh67}
Voshage, H. 1967, Z. Naturforschung, 22a, 477

\bibitem[{Voshage \& Feldmann(1979)}]{vosh79}
Voshage, H. \& Feldmann, H. 1979, Earth and Planetary Science Letters, 45, 293

\bibitem[{Voshage {et~al.}(1983)Voshage, Feldmann, \& Braun}]{vosh83}
Voshage, H., Feldmann, H., \& Braun, O. 1983, Z. Naturforschung, 38a, 273

\bibitem[{Wallmann(2004)}]{wall04}
Wallmann, K. 2004, Geochem., Geophys., Geosyst., 5, Q06004,
  doi:10.1029/2003GC000683

\bibitem[{Wasson \& Kallemeyn(2002)}]{wass02}
Wasson, J.~T. \& Kallemeyn, G.~W. 2002, Geochimica et Cosmochimica Acta, 66,
  2445

\end{thebibliography}
\bibliographystyle{aa}

{\small
\begin{longtable}{lllllllllll}
\caption{\label{tab:data}
Data base of meteorite CR exposure ages. Given are name of meteorite,
chemical group (original group in parentheses, see text), data source
(V79 for \citet{vosh79}, V83 for \citet{vosh83}), CR exposure age $t$,
error in exposure age $\sigma(t)$.  $t_\mathrm{100,+}$,
$\sigma(t_\mathrm{100},+)$, $t_\mathrm{100,-}$, and
$\sigma(t_\mathrm{100},-)$ are values after combining meteorites
within $\Delta(t)<100$~Myr of age, `+' combining intervals with
increasing age, `--' with decreasing ages. $t_\mathrm{H}$ and
$\sigma(t_\mathrm{H})$ correspond to values computed with the
hierarchical filter used in S02/S03. The triangles mark entries that
have been combined to the value given below (\tu) or above (\td),
respectively. Meteorites with suffix `-An' have an anomalous chemical
composition.
}\\
\hline\hline
Name& Group&Source&
$t$& $\sigma(t)$& 
$t_\mathrm{100,+}$& $\sigma(t_\mathrm{100,+})$&
$t_\mathrm{100,-}$& $\sigma(t_\mathrm{100,-})$&
$t_\mathrm{H}$& $\sigma(t_\mathrm{H})$\\
\hline
\endfirsthead
\caption{continued.}\\
\hline\hline
Name& Group&Source&
$t$& $\sigma(t)$& 
$t_\mathrm{100,+}$& $\sigma(t_\mathrm{100,+})$&
$t_\mathrm{100,-}$& $\sigma(t_\mathrm{100,-})$&
$t_\mathrm{H}$& $\sigma(t_\mathrm{H})$\\
\hline
\endhead
\hline
\endfoot
Morradal             &An            &V79 & 155  & 90 & 155& 90&   155&   90&   155&   90\\
South Byron          &An            &V79 & 255  & 70 & 255& 70&   255&   70&   255&   70\\
Washington County    &An            &V79 & 575  & 80 & 575& 80&   575&   80&   575&   80\\
Pinon                &An            &V79 & 790  & 50 & 790& 50&   790&   50&   790&   50\\
Deep Springs         &An            &V79 &2275  & 65 &2275& 65&  2275&   65&  2275&   65\\
Surprise Springs     &IAB (IA)      &V83 & 130  &170 & \tu&\tu&   135&  200&   134&   200\\
Bohumilitz           &IAB (IA)      &V79 & 140  &230 & 135&200&   \td&  \td&   \td&  \td\\
Rifle                &IAB (IA)      &V79 & 490  & 70 & \tu&\tu&   493&   77&   493&   75\\
Mayerthorpe          &IAB (IA)      &V79 & 495  &105 & \tu&\tu&   \td&  \td&   \td&  \td\\
Osseo                &IAB (IA)      &V79 & 495  & 55 & 493& 77&   \td&  \td&   \td&  \td\\
Canyon Diablo        &IAB (IA)      &V79 & 645  &103 & \tu&\tu&   648&   89&   648&   89\\
Bogou                &IAB (IA)      &V79 & 650  & 75 & 648& 89&   \td&  \td&   \td&  \td\\
Balfour Downs        &IAB (IA)      &V79 & 840  &110 & \tu&\tu&   840&  110&   902&   76\\
Odessa               &IAB (IA)      &V79 & 875  & 70 & \tu&\tu&   910&   66&   \td&  \td\\
Bischtuebe           &IAB (IA)      &V79 & 895  & 75 & \tu&\tu&   \td&  \td&   \td&  \td\\
Yardymly Aroos       &IAB (IA)      &V79 & 920  & 50 & 882& 76&   \td&  \td&   \td&  \td\\
Mount Ayliff         &IAB (IA)      &V79 & 950  & 70 & 950& 70&   \td&  \td&   \td&  \td\\
Deport               &IAB (IA)      &V79 &1140  & 70 &1140& 70&  1140&   70&  1140&   70\\
Nocoleche            &IC            &V79 & 250  & 70 & 250& 70&   250&   70&   250&   70\\
Bedego               &IC            &V79 & 940  & 90 & \tu&\tu&   948&   90&   947&   90\\
Arispe               &IC-An         &V79 & 955  & 90 & 948& 90&   \td&  \td&   \td&  \td\\
Smithonia            &IIAB (IIA)    &V79 &  90  & 80 & \tu&\tu&   144&   96&   142&   92\\
Sierra Gorda         &IIAB (IIA)    &V79 & 140  &110 & \tu&\tu&   \td&  \td&   \td&  \td\\
El Burro             &IIAB (IIB)    &V79 & 165  &115 & \tu&\tu&   \td&  \td&   \td&  \td\\
Cedartown            &IIAB (IIA)    &V79 & 180  & 80 & 144& 96&   \td&  \td&   \td&  \td\\
Lombard              &IIAB (IIA)    &V79 & 295  &200 & \tu&\tu&   325&  135&   339&  135\\
Sikhote Alin         &IIAB (IIB-An) &V79 & 355  & 70 & 325&135&   \td&  \td&   \td&  \td\\
Calico Rock          &IIAB (IIA)    &V83 & 545  & 55 & 545& 55&   545&   55&   545&   55\\
Sandia Mountains     &IIAB (IIB)    &V79 & 720  &160 & 720&160&   720&  160&   720&  160\\
Ainsworth            &IIAB (IIB)    &V79 &1280  &110 &1280&110&  1280&  110&  1280&  110\\
Wiley                &IIC-An        &V79 & 810  & 90 & 810& 90&   810&   90&   810&   90\\
Unter M\"assing      &IIC           &V83 &1385  & 70 &1385& 70&  1385&   70&  1385&   70\\
Brownfield           &IID           &V79 & 355  & 70 & 355& 70&   355&   70&   355&   70\\
Carbo                &IID           &V79 & 850  &140 & 850&140&   850&  140&   850&  140\\
Sacramento Mountains &IIIAB (IIIA)   &V79& 315  & 55 & 315& 55&   315&   55&   315&   55\\
Descubridora Charkas &IIIAB (IIIA)   &V79& 510  &110 & \tu&\tu&   548&  100&   510&  110\\
Sanderson            &IIIAB (IIIB)   &V79& 585  & 90 & \tu&\tu&   \td&  \td&   615&   78\\
Trenton              &IIIAB (IIIA)   &V79& 605  & 60 & 567& 87&   652&   73&   \td&  \td\\
San Angelo           &IIIAB (IIIA)   &V79& 610  & 80 & \tu&\tu&   \td&  \td&   \td&  \td\\
Tamarugal            &IIIAB (IIIA)   &V79& 610  & 85 & \tu&\tu&   \td&  \td&   \td&  \td\\
Treysa               &IIIAB (IIIB-An)&V79& 620  & 60 & \tu&\tu&   \td&  \td&   \td&  \td\\
Merceditas           &IIIAB (IIIA)   &V79& 625  & 80 & \tu&\tu&   \td&  \td&   \td&  \td\\
Picacho              &IIIAB (IIIA)   &V83& 635  & 50 & \tu&\tu&   \td&  \td&   \td&  \td\\
Lenarto              &IIIAB (IIIA)   &V79& 670  & 80 & \tu&\tu&   \td&  \td&   719&   63\\
Gundaring            &IIIAB (IIIA)   &V79& 685  & 90 & \tu&\tu&   \td&  \td&   \td&  \td\\
Joe Wright Mts       &IIIAB (IIIB)   &V83& 685  & 70 & \tu&\tu&   \td&  \td&   \td&  \td\\
Puende del Zacate    &IIIAB (IIIA)   &V79& 690  & 85 & \tu&\tu&   \td&  \td&   \td&  \td\\
Norfolk              &IIIAB (IIIA)   &V79& 695  & 67 & \tu&\tu&   \td&  \td&   \td&  \td\\
Grant                &IIIAB (IIIB)   &V79& 695  & 65 & 656& 74&   \td&  \td&   \td&  \td\\
Mount Edith          &IIIAB (IIIB)   &V79& 715  & 65 & \tu&\tu&   750&   61&   \td&  \td\\
Santa Apolonia       &IIIAB (IIIA)   &V79& 740  & 65 & \tu&\tu&   \td&  \td&   \td&  \td\\
Williamstown         &IIIAB (IIIA)   &V79& 740  & 55 & \tu&\tu&   \td&  \td&   \td&  \td\\
Thunda               &IIIAB (IIIA)   &V79& 755  & 60 & \tu&\tu&   \td&  \td&   \td&  \td\\
Delegate             &IIIAB (IIIB-An)&V79& 800  & 60 & 750& 61&   \td&  \td&   \td&  \td\\
Dayton               &IIICD (IIID)   &V79& 215  & 85 & 215& 85&   215&   85&   215&   85\\
Anoka                &IIICD (IIIC)   &V79& 600  &150 & \tu&\tu&   618&  110&   624&  110\\
Carlton              &IIICD (IIIC)   &V79& 635  & 70 & 618&110&   \td&  \td&   \td&  \td\\
Mundingi             &IIICD (IIIC)   &V79& 790  &100 & \tu&\tu&   792&   80&   793&   80\\
Edmonton (KY)       &IIICD (IIICD-An)&V83& 795  & 60 & 792& 80&   \td&  \td&   \td&  \td\\
Rhine Villa          &IIIE          &V83 & 325  & 70 & 325& 70&   325&   70&   325&   70\\
Kokstad              &IIIE          &V83 & 470  & 70 & \tu&\tu&   470&   70&   534&   72\\
Willow Creek         &IIIE          &V83 & 560  & 57 & 515& 64&   568&   74&   \td&  \td\\
Coopertown           &IIIE          &V83 & 575  & 90 & 575& 90&   \td&  \td&   \td&  \td\\
Nelson County        &IIIF          &V79 & 490  & 55 & 490& 55&   490&   55&   490&   55\\
Clark County         &IIIF          &V79 &1420  & 55 &1420& 55&  1420&   55&  1420&   55\\
Duchesne             &IVA           &V79 & 220  & 70 & \tu&\tu&   220&   70&   220&   70\\
Yanhuitlan           &IVA           &V79 & 300  & 65 & 260& 68&   342&   65&   355&   66\\
Seneca Township      &IVA           &V79 & 360  & 50 & \tu&\tu&   \td&  \td&   \td&  \td\\
Charlotte            &IVA           &V79 & 365  & 80 & \tu&\tu&   \td&  \td&   \td&  \td\\
Iron River           &IVA           &V79 & 400  & 70 & \tu&\tu&   448&   76&   \td&  \td\\
Putnam County        &IVA           &V79 & 435  & 70 & \tu&\tu&   \td&  \td&   461&   78\\
Huizopa              &IVA           &V79 & 450  & 90 & 402& 72&   \td&  \td&   \td&  \td\\
Hill City            &IVA           &V79 & 475  & 90 & \tu&\tu&   \td&  \td&   \td&  \td\\
Bristol              &IVA           &V79 & 480  & 60 & 478& 75&   \td&  \td&   \td&  \td\\
Maria Elena          &IVA           &V79 & 775  & 50 & 775& 50&   775&   50&   775&   50\\
Tawallah Valley      &IVB           &V79 & 250  & 85 & \tu&\tu&   250&   85&   250&   85\\
Hoba                 &IVB           &V79 & 340  &110 & 295& 98&   365&   80&   374&   80\\
Weaver Montains      &IVB           &V79 & 390  & 50 & 390& 50&   \td&  \td&   \td&  \td\\
Cape of Good Hope    &IVB           &V79 & 775  & 70 & 775& 70&   775&   70&   775&   70\\
Tlacotepec           &IVB           &V79 & 945  & 55 & \tu&\tu&   945&   72&   945&   73\\
Skookum Klondige     &IVB           &V79 & 945  & 90 & 945& 72&   \td&  \td&   \td&  \td\\
Glorieta Mountains   &PAL           &V79 & 230  & 70 & 230& 70&   230&   70&   230&   70\\
\hline
\end{longtable}
}

\end{document}